\documentclass[10pt]{article}

\usepackage{subcaption}
\usepackage{rotating}  %used to rotate things like pictures or things you \put down
\usepackage{theorem}   %for mathematical theorems and lemmas
\usepackage{bm}  %bold fond inside equations, eg. $\bm{\alpha}$
\usepackage{amsmath,amsfonts,amssymb,mathabx}
\usepackage{booktabs}  %Makes nice tables in books (sorta like RevTeX can)
\usepackage{pdfsync}   %used to sync pdf output inorder to flip from tex file to pdf and back
\usepackage{pdfpages}
\usepackage{float}
\usepackage{threeparttable}
\usepackage{makecell}
\usepackage{adjustbox}
\usepackage{listings}
\usepackage{algorithm}

\usepackage[explicit]{titlesec}
\usepackage{hyphenat}
\usepackage{ragged2e}
\RaggedRight

\usepackage{algorithmic}
\usepackage{multicol}

\usepackage{rotating}  %used to rotate things like pictures or things you \put down
\usepackage{theorem}   %for mathematical theorems and lemmas
\usepackage{bm}  %bold fond inside equations, eg. $\bm{\alpha}$
\usepackage{amsmath,amsfonts,amssymb}
\usepackage{booktabs}  %Makes nice tables in books (sorta like RevTeX can)
\usepackage{pdfsync}   %used to sync pdf output inorder to flip from tex file to pdf and back
\usepackage{graphicx}
\usepackage{latexsym}
\usepackage{amsmath}
\usepackage{amssymb}
\usepackage{fancyvrb}
\usepackage{colortbl}
\usepackage{enumerate}
\usepackage{pifont}
\usepackage{stmaryrd}
\usepackage{textcomp}
\usepackage{fncylab}
\usepackage{multirow}
\usepackage{paralist}
\usepackage{wrapfig}
\usepackage{longtable}
\usepackage{lscape}
\usepackage{hyperref}
\usepackage{lipsum}
\usepackage{garamondx}
\usepackage[T1]{fontenc}
\usepackage{amsmath}
\usepackage{graphicx}
\usepackage{xcolor}

\usepackage{geometry}
\usepackage{fancyhdr}
\usepackage[labelfont={footnotesize,bf} , textfont=footnotesize]{caption}
\makeatletter
\renewcommand\tagform@[1]{\maketag@@@ {\ignorespaces {\footnotesize{\textbf{Equation}}} #1.\unskip \@@italiccorr }}
\makeatother
\titleformat{\section}
  {\normalfont}{\thesection}{1em}{\MakeUppercase{\textbf{#1}}}
\titlespacing\section{0pt}{0pt}{-10pt}
\titleformat{\subsection}
  {\normalfont}{\thesubsection}{1em}{\textit{#1}}
\titlespacing\subsection{0pt}{0pt}{-8pt}

\makeatletter
\newcommand\sixteen{\@setfontsize\sixteen{17pt}{6}}
\renewcommand{\maketitle}{\bgroup\setlength{\parindent}{0pt}
\begin{flushleft}
\sixteen\bfseries \@title
\medskip
\end{flushleft}
\textit{\@author}
\egroup}
\makeatother

\usepackage[sort&compress]{natbib}
\makeatletter
\renewcommand\@biblabel[1]{\textbf{#1.}\hfill}
\makeatother

\bibpunct{}{}{,~}{s}{,}{,}
\colorlet{punct}{red!60!black}
\definecolor{background}{HTML}{EEEEEE}
\definecolor{delim}{RGB}{20,105,176}
\colorlet{numb}{magenta!60!black}

\lstdefinelanguage{json}{
    basicstyle=\normalfont\ttfamily,
    numbers=left,
    numberstyle=\scriptsize,
    stepnumber=1,
    numbersep=8pt,
    showstringspaces=false,
    breaklines=true,
    frame=lines,
    backgroundcolor=\color{background},
    literate=
     *{0}{{{\color{numb}0}}}{1}
      {1}{{{\color{numb}1}}}{1}
      {2}{{{\color{numb}2}}}{1}
      {3}{{{\color{numb}3}}}{1}
      {4}{{{\color{numb}4}}}{1}
      {5}{{{\color{numb}5}}}{1}
      {6}{{{\color{numb}6}}}{1}
      {7}{{{\color{numb}7}}}{1}
      {8}{{{\color{numb}8}}}{1}
      {9}{{{\color{numb}9}}}{1}
      {:}{{{\color{punct}{:}}}}{1}
      {,}{{{\color{punct}{,}}}}{1}
      {\{}{{{\color{delim}{\{}}}}{1}
      {\}}{{{\color{delim}{\}}}}}{1}
      {[}{{{\color{delim}{[}}}}{1}
      {]}{{{\color{delim}{]}}}}{1},
}

\title{Towards Global, Socio-Economic, and Culturally Aware Recommender Systems}

\author{
Kelley Ann Yohe*$^{a}$ \\ \medskip
$^{a}$Macquarie University, Sydney, Australia \\  \medskip
kelley.yohe@students.mq.edu.au
}

\begin{document}

% Makes the title and author information appear.
\vspace*{.01 in}
\maketitle
\vspace{.12 in}

% Abstracts are required.
\section*{abstract}

Recommender systems have gained increasing attention to personalise consumer preferences. While these systems have primarily focused on applications such as advertisement recommendations (e.g., Google), personalized suggestions (e.g., Netflix and Spotify), and retail selection (e.g., Amazon), there is potential for these systems to benefit from a more global, socio-economic, and culturally aware approach, particularly as companies seek to expand into diverse markets. This paper aims to investigate the potential of a recommender system that considers cultural identity and socio-economic factors. We review the most recent developments in recommender systems and explore the impact of cultural identity and socio-economic factors on consumer preferences. We then propose an ontology and approach for incorporating these factors into recommender systems. To illustrate the potential of our approach, we present a scenario in consumer subscription plan selection within the entertainment industry. We argue that existing recommender systems have limited ability to precisely understand user preferences due to a lack of awareness of socio-economic factors and cultural identity. They also fail to update recommendations in response to changing socio-economic conditions. We explore various machine learning models and develop a final artificial neural network model (ANN) that addresses this gap. We evaluate the effectiveness of socio-economic and culturally aware recommender systems across four dimensions: Precision, Accuracy, F1, and Recall. We find that a highly tuned ANN model incorporating domain-specific data, select cultural indices and relevant socio-economic factors predicts user preference in subscriptions with an accuracy of 95\%, a precision of 94\%, a F1 Score of 92\%, and a Recall of 90\%. \textbf{Keywords}: Recommender systems, personalization, cultural identity, socio-economic factors, user preferences.

% Keywords are required.
\section*{keywords}
Recommender systems, personalization, cultural identity, socio-economic factors, user preferences.

\vspace{.12 in}

% Start the main part of the manuscript here.
% Comment out section headings if inappropriate to your discipline.
% If you add additional section or subsection headings, use an asterisk * to avoid numbering.

%%%%%%%%%%%%%%%%%%%%%%%%%%%%%%%%%%%%%%%
\section{introduction}

Recommender systems have become an integral part of many consumer-related applications in recent years, providing personalised options to customers~\cite{elahi2021recommender,yakhchi2020towards}. These systems are primarily used by companies such as Google, Netflix, Amazon, and Spotify to recommend advertisements, media, and products to consumers \cite{Jannach2016,yakhchi2022convolutional}. However, these systems have been limited in their ability to incorporate socio-economic and cultural factors that can influence consumer preferences.
As companies expand into diverse markets, the need for more culturally aware and socio-economically sensitive recommender systems is becoming increasingly important. In this context, a recommender system that considers cultural identity and socio-economic factors have the potential to improve customer satisfaction and increase sales.

This paper aims to investigate the potential of a recommender system that considers cultural identity and socio-economic factors~\cite{beheshti2020towards}. Accordingly, in this paper, we will discuss the most recent developments in recommender systems and explore the impact of cultural identity and socio-economic factors on consumer preferences. We will propose an ontology and approach for incorporating these factors into recommender systems. Finally, we will present a scenario involving consumer subscription and plan selection in the media entertainment industry to illustrate the potential of our approach.

% In this template, an MQ colour %theme is used and the colours are provided in Figure \ref{MQcolours}.

%\begin{figure}[h]  
	%\begin{center}
	%\includegraphics[width=0.5\linewidth]{MQcolourtheme.png}
	%\end{center}
	%\caption{MQ colour theme}{This is the Macquarie University colour theme and is used for references and %hyperlinks in this Overleaf template.}
	%\label{MQcolours}
%\end{figure}

%Background
\subsection{Recommendations as a Competitive Advantage }

As technological advancements progress, an increasing amount of data is being generated through human activities, devices, and system processes. The introduction of openly available data has led to an overwhelming number of choices available to users, surpassing their capacity for understanding. To remain competitive and establish a strong relationship with their customers, businesses are required to assist users in sorting through this vast amount of information in order to uncover meaningful preferences~\cite{jalayer2022ham}. The ability to accurately identify preferences, provide recommendations, and profit from those recommendations represent a rapidly expanding and highly competitive business opportunity. Indeed, it can be a decisive factor in the success of a business strategy that leverages available customer knowledge to generate revenue, sustain profitability, and achieve a competitive edge.

Despite the potential benefits that recommender systems can offer, significant challenges are associated with their design and implementation. One major issue is the lack of global, socio-economic, and cultural awareness in current recommender systems, which can result in inaccurate recommendations and lost revenue opportunities.
For example, many current recommender systems rely heavily on users' personal data~\cite{beheshti2020personality2vec}, such as their search history, purchase behavior, and social media activity. However, this approach may not be effective in diverse markets where cultural and socio-economic factors significantly influence consumer behavior. In such cases, it may be necessary to develop recommender systems capable of considering these factors and adapting to local contexts.

In addition to the challenge of cultural and socio-economic awareness, there is also a need for transparency and ethical considerations in recommender systems. As recommender systems become increasingly integrated into our daily lives, there is a risk that they may perpetuate existing biases and inequalities, particularly in the case of sensitive issues such as healthcare, employment, and finance~\cite{rouhollahi2021towards}.
Despite these challenges, significant opportunities are also associated with developing global, socio-economic, and culturally aware recommender systems. By incorporating a wider range of factors into the design of recommender systems, it may be possible to provide more accurate and relevant recommendations to users, leading to increased customer satisfaction and loyalty. Moreover, businesses that can successfully navigate the challenges of global, socio-economic, and cultural awareness in their recommender systems may tap into new and previously under-served markets, leading to increased revenue and profitability.

\subsection{The Rationale for Intelligent Recommendations}

The advent of technology has resulted in an unprecedented increase in information generation, which has facilitated the development of more advanced systems with richer content. However, this expansion in content and system capabilities has also led to the potential for users to become overwhelmed by the sheer volume of available choices~\cite{sobhanmanesh2023cognitive}. This is particularly evident in the global connectivity that has been enabled by the internet, which has allowed for increased interconnections between societies, businesses, and individuals.
By 2025, worldwide data creation is expected to expand to a minimum of 180 zettabytes \footnote{https://www.statista.com/statistics/871513/worldwide-data-created/} through data connectivity over the internet, ultimately advancing globalization. E-commerce sites such as Amazon \footnote{https://www.amazon.com}, eBay \footnote{https://www.ebay.com}, Alibaba \footnote{https://www.alibaba.com}, etc. aggregate both products and businesses globally to create large catalogs of information available to a consumer.  Spotify \footnote{https://www.spotify.com} and Netflix \footnote{https://www.netflix.com} similarly create global catalogs of media, only restricted by licensing and user preferences driven by cultural and language preferences.

Humans, unlike machines, have inherent limitations in storing, retrieving, and processing large information~\cite{sensemaking}. This limitation has led to the development of mechanisms that augment human ability through more efficient filtering, sorting, summarizing, visualizing, and retrieving information~\cite{sensemaking}. However, the need for efficient information processing has become increasingly important in today's digital age, where the volume of data generated is staggering and ever-increasing. To address this challenge, various additional techniques have emerged, including the development of recommender systems~\cite{tabebordbar2019}.
Recommender systems are intelligent algorithms that assist users in decision-making processes by providing recommendations based on their preferences and past behavior. These systems have found widespread use in various industries, including e-commerce, music, and entertainment, and have proven to be effective in improving customer satisfaction and loyalty. By leveraging machine learning and data mining techniques, recommender systems can analyze user data and make personalized recommendations, thus reducing the information overload that users may experience~\cite{wang2023learning}.
Despite their potential benefits, recommender systems are not without their limitations. One challenge is the difficulty in accurately modeling user preferences, particularly in cases where cultural and socio-economic factors play a significant role in decision-making. Accordingly, it is essential to explore the development of more sophisticated recommender systems that incorporate a wider range of factors, including cultural and socio-economic considerations. Such systems would require a greater understanding of human decision-making processes and the factors that influence them and would need to be designed with transparency and ethical considerations in mind.

\subsection{Globalization and Plan Selection in Media}

Globalization is described as "how trade and technology have made the world into a more connected and interdependent place"~\cite{NatGeographicEnc}. The digital and information era has accelerated globalization. This phenomenon of accelerated globalization has brought tremendous business opportunities for various industries, including the media and entertainment industry. With the advent of the internet and information and communication technologies, businesses can now expand into previously restricted markets due to language, regulations, barriers in product delivery, and commercial costs~\cite{Kraemer2005}. The media entertainment industry, in particular, has traditionally been localized, owing to constraints in localized linear TV programming. Linear TV programming is delivered via a network provider, where the provider has predefined programming, ads, and time segments~\cite{LinearTV2021Aug}. However, with the emergence of streaming media and on-demand programming, which account for a minimum of 63\% of consumer viewing~\cite{LinearTV2021Aug}, linear TV has faced intense competition. In response, companies such as Netflix, DisneyPlus + Hulu, and AmazonPrimeVideo have recognized the profitability and availability of the global market within the last decade~\cite{NetflixGlobal2016}, leading to increased competition for market share.

While globalization presents profitability opportunities, it also poses several challenges in business strategy~\cite{Caprar2015}. To profit from globalization, businesses need to strike a balance between streamlining product operations and tailoring products to meet consumer expectations, which may differ from country to country. Differences in user preferences due to socio-economic, geographic, or cultural variations between countries and regions can create segmentation on product preferences and pricing~\cite{Bolton2003}. This complexity is often apparent in product-based media subscription plans and their prices.

In light of these challenges, businesses in the media entertainment industry need to develop and adopt innovative approaches that enable them to meet consumer expectations and capitalize on the available opportunities. Recommender systems have emerged as one such approach to assist users in their decision-making by providing intelligent recommendations. These systems have been widely used in the e-commerce, music, and entertainment industries~\cite{tabebordbar2019}. However, existing recommender systems often fail to consider the socio-economic, geographic, or cultural variations between countries and regions, thereby limiting their ability to understand user preferences~\cite{Shi2014} accurately. This limitation underscores the need for a more global, socio-economic, and culturally aware recommender system that can address the challenges of globalization and provide businesses with a competitive advantage.

\subsection{Problem Statement and Challenge}

Culture plays an important role in shaping an individual's preferences in a wide range of domains such as product selection, social interactions, education, and entertainment. Recommender systems have been extensively studied for personalizing recommendations in media, music, and consumer products. Previous studies have also evaluated the efficacy of these systems in providing recommendations that align with cultural, socioeconomic, and geographic contexts. Despite these efforts, the design of recommender systems that intentionally incorporate contextual information related to users' cultural, socioeconomic, and geographic backgrounds remains limited. This represents a significant research gap, as the lack of culturally, socioeconomically, and geographically aware recommendation systems may lead to ineffective or biased recommendations that do not adequately account for the specific needs and preferences of users from different backgrounds. Therefore, there is a need for further research that focuses on developing more inclusive and culturally sensitive recommender systems, which can effectively serve diverse user populations by accounting for the contextual factors that influence their preferences.

\subsection{Contribution}
This paper makes a significant contribution to the field of recommender systems by exploring the potential of a global, socio-economic, and culturally aware approach. The study's main focus is to understand the impact of cultural identity and socio-economic factors on consumer preferences and recommenders' accuracy. The proposed ontology and approach for incorporating these factors into recommender systems provide a basis for further research on this topic.
Moreover, the paper provides an illustration of the potential of the proposed approach by presenting a scenario of consumer subscription and plan selection in the media entertainment industry. The scenario demonstrates how current recommender systems fall short of understanding user preferences accurately due to their lack of awareness of socio-economic and cultural factors. The proposed approach addresses this limitation by updating recommendations in response to changing socio-economic conditions.

This research is significant because it highlights the need for a more comprehensive approach to recommender systems that can incorporate socio-economic and cultural factors. It contributes to the development of a more accurate and effective recommender system that can cater to diverse markets, and ultimately benefit companies seeking to expand globally.
In particular, this study presents a comprehensive analysis of the current state-of-the-art recommender systems, focusing on context-aware systems. Specifically, we investigate the potential of incorporating cultural, socio-economic, and geographic indicators as context into recommender systems. We begin by reviewing existing literature on cultural models and the limited research that has been conducted on incorporating these indicators into recommender systems. To identify relevant indicators, we leverage publicly available data on these factors for individual countries across the globe. Based on this analysis, we propose an ontology that incorporates these indicators and outlines the principal features that can be integrated into recommender systems to enhance the accuracy of recommendations.

\subsection{Summary and Outline}
The field of recommender systems has gained significant attention in recent years due to the increasing demand for personalized and customized options for customers. However, despite their popularity and widespread use, these systems have been criticized for their limitations in accurately understanding user preferences. Specifically, they lack awareness of socio-economic and cultural factors, which have a significant impact on consumer behaviour and preferences. Moreover, these systems often fail to update recommendations in response to changing socio-economic conditions. To address these limitations, this paper aims to investigate the potential of a recommender system that takes into account socio-economic and cultural factors. 

This section introduced the topic, providing a brief overview of the research problem and our approach to addressing it. We presented a motivating scenario and outlined the main objectives of our research. 
The subsequent sections of this paper are structured as follows:

Section 2 - Related Works:
This section presents a review of existing research on recommender systems. We will discuss the current state-of-the-art in various types of recommender systems and their underlying design principles for generating recommendations. We will devote additional attention to context-aware recommender systems, which are particularly relevant to our motivating scenario. Moreover, we will examine two primary models for assessing culture across different regions and countries. We will also survey the existing literature on how socio-economic and cultural contexts have been incorporated or evaluated in recommender systems.

Section 3 - Methods and Materials:
This section will outline the methodology employed to evaluate the efficacy of incorporating socio-economic and cultural indicators as context in recommender systems. Our proposed solution entails concatenating non-domain-specific and domain-specific indicators, which are subsequently standardized, classified, and modelled. The feature set is subjected to rigorous analysis to ensure optimal performance.

Section 4 - Experimentation Results and Evaluation:
This section will present the results obtained through the methodology described earlier. We will discuss the techniques utilized for feature engineering~\cite{khadivizand2020towards}, model comparison, and evaluation to understand the influence of socio-economic and cultural factors on predicting subscription preferences based on price.

Section 5 - Conclusion:
This section will represent the concluding section of our research and will provide a review of the achieved results and their effectiveness in meeting the intended research goals. In this section, we also briefly discuss potential areas of future research extension that the author intends to pursue.

%%%%%%%%%%%%%%%%%%%%%%%%%%%%%%%%%%%%%%%%%
\section{Related Work}

Recommender systems are systems such as those seen in Pandora, Netflix, and Spotify, where the system intelligently learns specific preferences for a user over time. Data presents itself as a user interacts with the system, and the system learns the user’s behaviour. As a result, the system suggests more and more accurate and intelligent recommendations to a user based on learned preferences through data and time. This data improves the decision-making process and eliminates information that may not be relevant. Recommender systems are used in various applications, including those most commonly found in media, social networking, and advertisement. In the case of advertising, recommender systems may also attempt to persuade a user into a purchase~\cite{Scheel2014}.

\subsection{Variations in Producing Recommendations} %Variations in Recommender Systems
With the maturity of Recommender Systems, several techniques have emerged in forming user preferences to produce a recommendation. These techniques aim to reduce information retrieval, distill information into tasks and goals, and finally, project a proactive prediction as a recommendation to a user. Various techniques are utilized to take on obstacles in certain domains. For instance, in the media domain, user feedback on previously made selections, such as liked or not liked, is captured and applied. These previously used selections, with additive user context, help narrow the future information space by applying a learned understanding of a user’s preference over time to similar type items, otherwise known as \textit{item-based} recommendations~\cite{YakhchiShahpar2021LCUP}. In this case, users may be grouped together based on their preferences. For example, a user searching for ski apparel may be similarly grouped with those searching for other snow apparel, such as down jackets. While all these techniques may vary, they all seek to narrow the information space, derive the intent of a specific user, and assist the user in selecting information, an experience, or a product relevant to either their need or the search.  
He et al.~\cite{he2023noise} proposed a Noise-augmented Contrastive Learning for Sequential Recommendation (NCL4Rec) that addresses the limitations of current methods by introducing noise-guided data augmentation and a unique noise loss function, leading to improved performance over state-of-the-art models in handling noisy data in sequential recommendations.
There is a minimum of four classes of recommender systems: Collaborative Filtering, Content-Based, Knowledge-Based, and Context Aware~\cite{Scheel2014}. 
We will explore these further in the upcoming sections.

\subsubsection{Collaborative Filtering Based Recommender}
Collaborative Filtering~\cite{Scheel2014}is a class of recommender systems dependent on the user’s interactions or feedback with the system. They are considered user-based systems. The system can either explicitly inquire about interactions or user feedback, or implicitly derive them. Explicit feedback is the feedback that is collected purposefully. Examples of explicit feedback may be such as those with likes or dislikes in song titles, or those of user ratings when reviewing restaurants. However, the system can also implicitly derive and collect feedback. Examples of implicit feedback are those such as when a user searches for a particular movie or book, or even the frequency by which a user may search for certain clothing. 

This feedback is then used to determine similarities or dissimilarities across users. Within collaborative filtering, we often compute similarity to determine as the method or likelihood of a user selecting a different item for similar interests or reasoning. The similarity is computed as a mechanism to describe implicit likeness between users based on their implicit and explicit feedback. In the example of music, these users have selected music titles similar to other users’ preferences or to other similar music titles they others have listened to in either the past or present (e.g. current music trends). We may consider these similar users in the same neighbourhood. These neighbours may have similarity computed based on either the \textit{other users’} likes or ratings, or based on the \textit{user’s own} ratings or \textit{likes}~\cite{Ricci2022}. The similarity relationship may be inferred based on either direction. Collaborative filtering is a flexible approach to variations in a user’s preference, where it is not dependent on the inference of metadata on an object, such as genre within the music example. The methods of calculating neighbours are based on machine learning techniques. They may use neighbourhood models, decision tree models, rule-based models~\cite{shabani2023rule}, latent factorization models, or even models such as Naive Bayes. 

%<potential example of users influence vs ratings-influenced filters>

\subsubsection{Content Based Recommender}
Content-based recommender systems~\cite{Aggarwal2016} are systems that heavily depend on metadata, data embedded or stored, or data that is systematically collected. This metadata is used to describe the attributes of an object or item. These recommender systems are typically item-based systems vs user-based systems. Similar to collaborative filtering in these systems, the recommendation outcome is computing similarity. However, in this case, the similarity is within the context of the user's liked or rated item. This metadata is data embedded or stored that describes an object or item used to find items uniquely similar to the item a user has previously liked or rated. Since similarity is item-based, content-based recommender systems almost exclusively focus on the owner’s rating or like as a key input to determining which items a similarity algorithm should initially be based on, vs a neighbouring user. Typically, in these systems, other users have no role. 

Attributes used in content-based recommender systems may either exist on the item or the user. In these cases, the metadata may be item data that is intentionally stored with the item, such as with a book, song, or movie. In the example of a book, the genre, the author, the publishing house, and other descriptive attributes are all stored with the book. A movie’s attributes would include information such as genre, actors, ratings, and other descriptors that may exist. A user’s attribute information may include domain-specific knowledge relative to that user. In the music example, this may include titles they have played, or recently played, the items they have liked or not liked, or inferred data, such as keyword searches, search history, or other similar examples. 

The steps to creating a content-based recommender system involve 1) item and domain-based feature extraction, 2) constructing a learning-based User Profile model, and lastly, 3) a method for filtering and predicting a recommendation. Feature extraction plays a pivotal role in content-based systems. We need to pre-process most of the attributes associated with the user or the item, often extracting them from the text. The emphasis on feature extraction is to provide specific areas of focus that have a meaningful impact on predicting the user’s interests. The features are then ranked or measured statistically for their impact. The learning-based user model provides a set of key characteristics that are learned and adapted based on a user’s previous history with items and their ratings, likes, or interactions. As new items of interest are determined, this model would adapt and change to reflect the user’s interests accordingly. Lastly, the user model and the features are used to make an effective recommendation for the user’s future interest.

\subsubsection{Knowledge Based Recommender}

Knowledge-based recommender systems~\cite{Rosa2019}, unlike collaborative filtering or content-based systems, heavily depend on interactions with the user. The user supplies the system with a set of requirements, then the system sources domain-specific data and combines the domain-specific data with item-specific attributes. Once complete, the system then predicts a recommendation. The process is typically iterative and allows users to further refine their requirements based on the set of information predicted. These systems require the use of domain-specific databases as a critical input to predicting the user’s preference. 

In previously described recommender systems, the attributes of items may be used independently or cooperatively to predict a user’s recommendation. Knowledge-based systems cooperatively use all of an item's attributes and match them with the user-provided requirements to make a prediction. A user may specify requirements that are constraint-based, such as those that may be associated with recommender systems predicting purchase decisions. Examples in this space may be those associated with flight, real estate, or vehicle purchase recommendations. These systems often have a predefined set of choices. A flight-based recommender system will request departure and arrival date and time information, preferences in stops, tolerance in the duration of layovers, alliance preference, and other such constraints. Default requirements may be used as a manner by which to guide a user to make a more refined selection. 

This refinement narrows results. Flights default to exactly that day, and users may also expand the search to +/- 3 days to broaden the results. A ranking process is also used to provide a way to simplify the presentation of information retrieved to the user. In these cases, the ranking process may be a single attribute. We may see an example of this within real estate, where a constraint in an area is drawn with boundaries in a user interactive map or a zip code. A utility function may frequently define a constraint. This utility function matches the attributes to a weighting that is based on a pre-determined assumption. The assumption is based on the utility of the attribute in reference to the user’s selection. For instance, within the real estate system, the system may specify that price or affordability may have a higher weight in displaying housing when matching a user preference. This system may set this weighting over a user’s prefer’s preference in the number of bathrooms if in fact, bathrooms were not specified by the user.   

A new type of recommender systems, called Cognitive Recommender Systems~\cite{beheshti2020towards}, focus on:
(i)~organizing the big data in a Data Lake~\cite{beheshti2017coredb}; 
(ii)~automatically contextualize it in a knowledge lake~\cite{beheshti2018corekg} to form a large knowledge graph;
(iii)~mimicking the knowledge of subject-matter experts (using crowdsourcing services) to enrich the knowledge grapg~\cite{beheshti2022knowledge,beheshti2023empowering}; 
(iv)~leverage the personality, behaviour, and attitude of users to enrich and augment the knowledge graph~\cite{beheshti2020personality2vec,liu2022dagad,barukh2021cognitive}; and
(v)~use novel embeddings approach to summarize the large knowledge graph~\cite{wang2023learning,shabani2023survey}.
The whole model is represented as ProcessGPT~\cite{beheshti2023processgpt}, which fed the enriched knowledge graph into an AI engine to enable intelligent recommendations.

\subsubsection{Context Aware Based Recommender} 

To define context-aware recommender systems, we must first explore context. This includes what it is, how it is defined in computing, what context awareness is, and how context manifests in systems. 
Scholars have described context in various ways in years past, so much so that its use in computing has been for numerous years hotly debated. Within the realm of research, the scope of how wide and how far context has been used was often defined or constrained by the researcher himself~\cite{Mowafi2014}. Context, however, has a clear purpose. That purpose is to provide clarity and efficiency in communication~\cite{PatrickBrezillon2014}. When the context is commonly shared across systems or users, it creates a basis for understanding. This understanding bridges cultural divides, traverses differences in language, and brings forth shared understanding in domain knowledge, either in breadth and depth, or cross-domain. Conversely, when the context does not exist between parties, communication may falter, creating misunderstanding. While many definitions have been used over the years, context may be thought of as providing reasoning, intent, or rationale to an event, occurrence, or thought such that it may be understood to the fullest capacity. The basis for this reasoning, intent, or rationale is often held within the characteristics surrounding an entity or user’s situation~\cite{Moradizeyveh2022intent}. 

Context awareness in computing is a phrase used to describe the state in which a system is situationally aware. Gartner uses the following definition to describe context-aware computing: \textit{"a style of computing in which situational and environmental information about people, places and things is used to anticipate immediate needs and proactively offer enriched, situation-aware and usable content, functions and experiences."}\footnote{https://www.gartner.com/en/information-technology/glossary/context-aware-computing-2}
In other words, context awareness primarily concerns the ability to \textit{sense} a change in either situation or environmental information and respond accordingly~\cite{AlanColman2014}. 
There are several ways in which we can define attributes that are representative of systems that exhibit context awareness. We can minimally define these attributes within a system as one of the following~\cite{Mowafi2014}:

\begin{enumerate}

    \item the awareness of the situational and environmental context surrounding an entity’s environment

    \item the ability to utilize awareness to personalize and make a recommendation that is intended to meet an entity’s interest through the context collected

    \item the ability to discover or sense, and adapt to the changing needs of the entity and their surrounding environment or situation and then revise a personalization and recommendation based on that discovery

    \item the ability for a system to take ambient signals, such as signals from IOT devices, mobile or biometric devices, to understand and appropriately inject into a recommendation whereby momentary adaptations frequently occur to meet the transient nature of a user

\end{enumerate}

Context-aware recommender systems~\cite{Scheel2014}, \cite{Aggarwal2016} are systems that often describe and make recommendations based on the relationship of the user and the item at the point of \textit{when} and \textit{where} the user is. They are designed on the presumption that the time (\textit{when}) and location (\textit{where}) of the user, at the point of selection, influence what will result in a successful recommendation by the system. In this case, the user's when and where are considered the \textit{context} in which a user is currently operating. This context can be described in physical conditions of when and where, and in non-physical conditions such as social, emotional, or activity-based conditions~\cite{BenSassi2017}. 
The following definitions can be used to understand the categories of context-aware systems~\cite{BenSassi2017}:

\begin{itemize}

\itemsep0em

\item \textbf{Location-Based} describes geographic attributes that describe where the user is.

\item \textbf{Time-Based (Temporal)} describes the point in time where the user exists.

\item \textbf{Social-Based} describes the social attributes of the user, such as their social network, social interests, etc.

\item \textbf{Emotion-Based} describes the feelings or emotions of a user.

\item \textbf{Activity-Based} describes what events or activities the user is engaged in. Event extraction approaches~\cite{li2021comprehensive,li2022survey} can be used for promptly apprehending event information from massive textual data.

\item \textbf{Multi-Dimension} describes the combination of any types of context that correlate with each other to form deeper situational awareness.

\end{itemize}

\begin{figure}[t!]

    \centering

    \includegraphics[width=1\textwidth, angle=0]{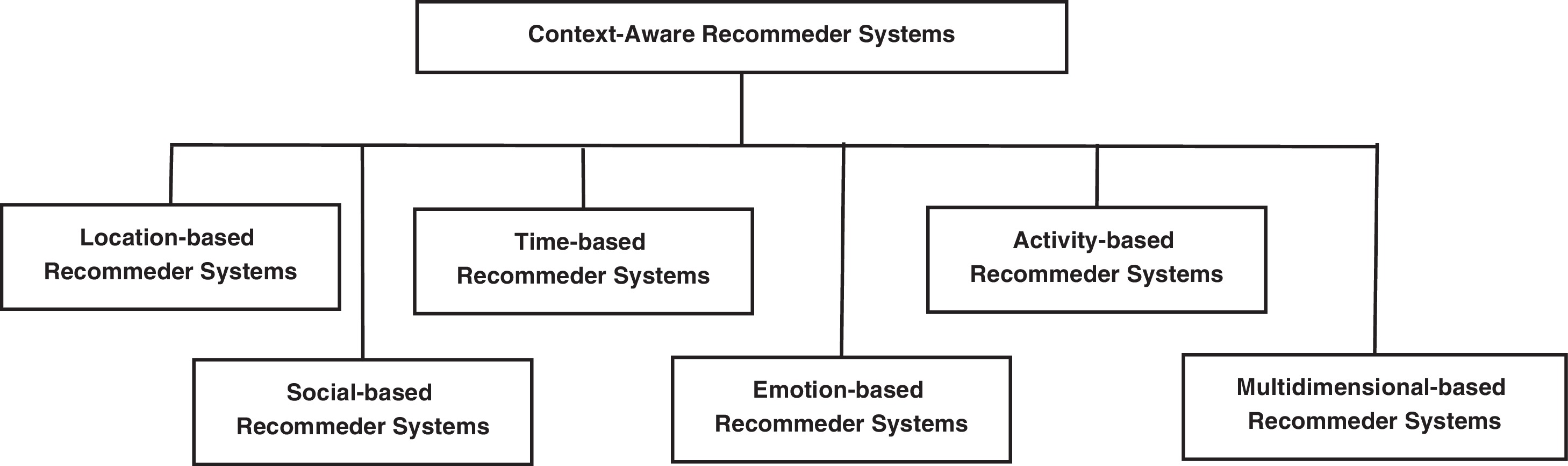}

    \caption{Various Types of Context Aware Systems \cite{BenSassi2017}}

    \label{fig:Context Aware Recommendation Process}

\end{figure}

Any number of these areas of conditions can be correlated and used to describe a multi-dimensional~\cite{BenSassi2017} approach to context. This multi-dimensional approach could be used in recommending certain types of music based on location, geography, and time of the year, e.g., where a recommendation combines Spanish influences within Christmas music because a user’s context is determined to be the Christmas holiday while on vacation in Mexico. 
Mathematically, the single and multi-dimensional cases are represented by, the item [I] and the possible recommendations [R], which are further developed by a set of [C] contextualized attributes~\cite{Adomavicius2005}. Types of context may describe the attributes of a specific location or time. For instance, with location, where location is a country, other types of supporting information, such as country-specific demographic information, may further describe the context of the user’s situational environment. This specific country’s demographic information may include language, currency, population, macroeconomics, and cultural or social characteristics. With advertisement-based recommender systems, a recommendation for an advertisement may include a country-specific version of an item. There are cases where the context of a user is used to identify similar users (e.g. users within the same geographic area) to provide a recommendation based on an already-learned understanding of other users. In these cases, unsupervised machine learning clustering techniques, such as K-Means, are used to group and then identify the most similar user(s). The attributes of the cluster in the geography example are those attributes adjacent in relation to geography, e.g. IP Address, longitude, and latitude. The Euclidean Distance~\ref{eq:2.1}, the line between two points on a plane, is often used to calculate the distance between users within the cluster. This helps to determine relationship proximity. This example is one explored in research on Cultural Heritage~\cite{Hong2019}.

\begin{equation}
    d\left( p,q\right)=\sqrt{\sum _{i=1}^{n}\left( q_{i}-p_{i}\right)^2} 
\end{equation}
\label{eq:2.1} 

2.1 Euclidian Distance. 
Euclidian Distance

The relationship proximity is then used to find the nearest neighbour \begin{math} \mathbf{q } \end{math}~in the cluster to our user~\begin{math} \mathbf{ p} \end{math} by which to inform a recommendation.
In some cases of measuring \textit{geographic distances} with more accuracy, the countries representing the users within a cluster are measured  by researchers utilizing Vincenty’s method of calculating the distance between countries~\cite{Hooijberg2007}. This method measures the shortest spherical distance between two points, where each point represents a geographic location of a single country on the earth's surface.

\begin{figure}[ht!]

    \centering

    \includegraphics[width=1\textwidth, angle=0]{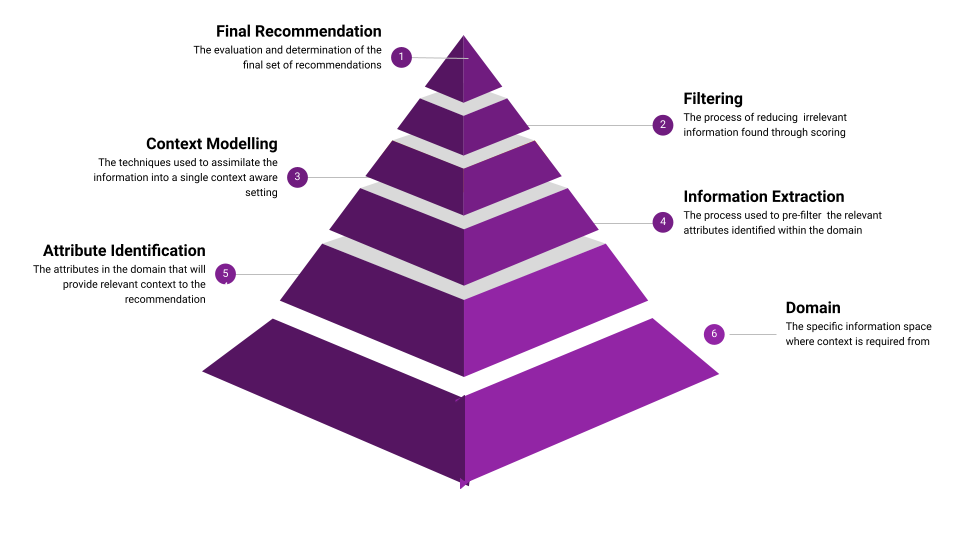}

    \caption{Processes in determining contextual recommendations} %some differences in the input of the networks, dimension of vector, and how to link the objects in 2 models.

    \label{fig:Context Aware Recommendation Process}

\end{figure}

\textbf{Information Processing for Context-Aware Recommendations}.
Context-aware recommender systems exist in many application domains. Researchers in 2017 generalized these domains into six areas: (i) travel and tourism, (ii) places, (iii) e-documents, (iv) multimedia, (v) e-commerce, and (vi) others~\cite{Haruna2017}.
To enrich a recommendation [R] with the right context [C], there are several stages to processing the set of possibilities to present to a user Figure~\ref{fig:Context Aware Recommendation Process}.  
The domain (6 - Domain) is the narrower set of information spaces that are used to extract context. Identifying the domain or space of context in which enrichment is targeted provides a base for the system. Within each domain, a process then occurs to determine the attributes (5 - Attribute Identification) of the domain that apply to the item [I]. Once these attributes are identified, the data related to these attributes are extracted from the domain (4 - Information Extraction). 
This extraction is a type of pre-filtering. Post-pre-filtering, the information is assimilated into a single model (3 - Context Modelling). There are several modelling techniques in Table ~\ref{tab:context-modeling-techniques}: Markup, Logic, Ontology, Vector, and Graph.  There is a final iterative filtering process (2 - Filtering), that reduces irrelevant information, often through scoring once a model is created, and then lastly a final recommendation is produced (1 - Final Recommendation).
% Please add the following required packages to your document preamble:
% \usepackage{booktabs}

\begin{table}[]
\centering
\begin{tabular}{@{}|l|l|ll@{}}
\cmidrule(r){1-2}
\textbf{Markup} &
  \textit{\begin{tabular}[c]{@{}l@{}}refers to the tagging, labelling, or inscription of metadata\\ to correlate context further.\end{tabular}} &
   &
   \\ \cmidrule(r){1-2}
\textbf{Logic} &
  \textit{\begin{tabular}[c]{@{}l@{}}refers to the application of logic as a way of describing the\\ correlation of context with the item or user\end{tabular}} &
   &
   \\ \cmidrule(r){1-2}
\textbf{Ontology} & \textit{\begin{tabular}[c]{@{}l@{}}describes the formalization of context into an organized\\  structure\end{tabular}} &  &  \\ \cmidrule(r){1-2}
\textbf{Vector}   & \textit{describes the mathematical representation of  information}                                                     &  &  \\ \cmidrule(r){1-2}
\textbf{Graph}    & \textit{\begin{tabular}[c]{@{}l@{}}refers to the organization of information based on \\ relationship\end{tabular}}    &  &  \\ \cmidrule(r){1-2}
\end{tabular}
\caption{Techniques in modeling context \cite{Haruna2017}}
\label{tab:context-modeling-techniques}
\end{table}

\textbf{Temporal Drifts in Context-Aware Recommender Systems}.
In recommendations, several factors may be considered when formulating a user’s preference. To predict user preferences, different models have been used with rating scales to produce relatively static predictions. Humans, however, are more fluid in their preferences, influenced by others, and develop and change preferences over time. In its simplest form, these systems use rating scales. However, to consider recommendations more accurately, the system must consider the human biases that may occur over time. These biases could be due to cultural influences, economic influences, social influences, or rare, but occasional, personality shifts. An example of a change in preference over time may be a user’s preference towards animated movies when they are younger, vs action movies as they mature towards a young adulthood. These instances of preferences in time are called temporal drifts~\cite{Zafari2019}. Temporal drifts can be modelled to show how changes affect the domain or a user. Zafari~\cite{Zafari2019} modelled temporal changes that were tested with both social influences and conditional preferences at various periods in time. In his work, Zafari included preferences arising from social influences for certain feature categories. These were tested in very domain-specific areas, such as music. 

In these domain-specific areas categories, entity biases were modelled using latent factor techniques to consider changes in preferences. Latent factorization, a mathematical approach to recognizing otherwise less observed factors in a model, was used to provide a heavier weighting to recently selected preferences by a user. Koren termed this approach to temporal-based recommender systems, Latent Temporal Based Optimizations (LTO)~\cite{Koren2009}. Further work has been done to include additional considerations around the dynamic nature of certain categories of preferences, and the ability to capture the granularity of detail within these categories more specifically. This granularity allowed for the specificity needed to highlight static features that were unchanged due to temporal drift vs features that were highly likely to be influenced by temporal drift~\cite{Koenigstein2011}.
Other research reviewed variations in approaches to determine temporal drift using Bayesian models, tensor factorization, and memory-based time-aware systems. While these approaches in modelling differed, several of them added a weighting approach to the preferences to mature the dynamic aspects of features that were heavily influenced by temporal drift~\cite{Zafari2019}.

\subsection{Culture and Culture Models}

Culture can be described as “the collective programming of the mind which distinguishes the members of one human group from another”~\cite{Hofstede2011}.
Over the years, several models of culture have been developed and used to both evaluate and understand various domains and the effects of culture in models.  These various models have been applied in research to differentiate cultures and provide the relevant context where culture is a factor. There are a few prominent models in particular, one by Hofstede, and another by Tropenaar.

\subsubsection{Hofstede’s Cultural Model}

Hofstede developed a well-regarded model for evaluating culture across several key dimensions. Each dimension is assigned pre-computed indices that describe each country through six distinct measures.  These measures can be used to describe the spatial proximity between countries on each dimension. This relativity between indices is used to evaluate comparability and similarity across countries both individually at each dimension and in totality across all six dimensions. The six dimensions~\cite{Hofstede2011} that Hofstede defined are: 

\begin{enumerate}

    \item  \textbf{Power Distance} - the degree at which inequality within society is expected and respected or not expected and disrespected. For example, high degrees of power distance is associated with centralized or hierarchical societies.

    \item \textbf{Individualism} - this is the importance that individuals place on the need to have collective structures and interdependence, or autonomy and self. 

    \item \textbf{Masculinity} - represents a preference for gender-dominant traits within a society. These often describe how society has an inherent need for achievement, assertion, heroism, competition, or material rewards for success. Societies largely more consensus-based, meek, modest, cooperative, and ethereal are regarded as femininity. 

    \item \textbf{Uncertainty Avoidance} - represents the most dominant behaviours in how a society thinks about and prepares for the future and how comfortable individuals within the society are with ambiguity.

    \item \textbf{Long Term Orientation} - understands how a societal group may perceive and make decisions in relation to current events pragmatically, or seek to retain its past and traditional norms.  

    \item \textbf{Indulgence} - describes the society’s tolerance or orientation towards innate human needs that directly relate to enjoyment vs an orientation towards alignment with social norms that suppress innate needs related to enjoyment.

\end{enumerate}

\subsubsection{Tropenaar’s Cultural Model}

Other cultural models have also been developed, showing slightly different models, but representing extremes in measures to quantify differences accurately. For instance, Tropenaar~\cite{Trompenaars1998}, also ascertained in his research that the more we could understand cultural differences, the more we would be able to bridge those differences in social, relational, community, government, and work. He defined cultural differences around the differences between seven different pairs of attributes:  1) universalistic and particularistic 2) individualistic and communitarian 3) specific and diffuse 4) neutral and affective 5) achieved and ascribed, 6) sequential and synchronic 7) internal and external. These can be explained in more detail below.

\begin{enumerate}

    \item  \textbf{Universalistic vs Particularistic} - Universalistic describes the emphasis in one culture to put universal norms and truths over relational or achievement-based ones.  In Particularistic, relational, and achievement-based cultures, laws and norms exist to help define the interaction between people and how they relate. 

    \item   \textbf{Individualistic and Communitarian} - Communitarian cultures are cultures that emphasize and reward the needs of the community over the needs of the individual. In these cultures, individuals are expected to put the community's needs over their own. Individualistic cultures emphasize the needs, happiness, and welfare or fulfilment of the individual, and their systems reward such.

    \item \textbf{Specific vs Diffuse}  - Specific cultures are those that can differentiate between each component and recognize how they form a larger structure. These cultures use facts, measures, standards, and contracts to define individual components. Business, for instance, is referred to and understood based on these facts~\cite{beheshti2016business}. Diffused cultures, however, rely on relationships and uphold values of empathy, trust, and reliability~\cite{schiliro2022deepcog}. The significance of these relationships is greater than that of facts, measures, standards, and contracts. Achieving business aims in this culture begins with creating relationships.

    \item \textbf{Affective vs Neutral} - These cultural attributes describe the degree to which the display of emotions by an individual is acceptable. Neutral cultures deem showing emotion to diminish power and influence. These cultures encourage the curtailing of emotions.

    \item \textbf{Achieved vs Ascribed} - These cultural attributes describe how individuals within a culture are assigned ‘status’. In achieved cultures, status is derived from results (achievements). An individual’s ability to maintain their status is derived from the continuity of results. In ascribed cultures, status is assigned by class, birth, rank, age, gender, or wealth.

    \item \textbf{Sequential vs Synchronic} - Time and how people within a culture think of time can be a differentiator. Sequential cultures value planning.  They value the sequence of plans, and keeping to defined plans over adapting. Synchronic cultures value the ability to execute time-based events in parallel and adjust as needed. 

    \item \textbf{Internally or Externally Controlled} - The relationship with the natural environment is also a cultural identifier. Internally controlled societies believe they can exert and direct their lives, including over natural forces within the environment. Externally controlled societies believe they must exist in harmony with their natural environment. They exert limited or no control over how their destiny is shaped.
\end{enumerate}

\subsection{Socio-economic and Cultural Influences in Context-Aware Recommender Systems}

As the use of recommender systems is still early, there has been a limited amount of research reviewing the outcomes of either socio-economic, or cultural influences in context-aware recommender systems, and even more limited is the combination of both.

\subsubsection{Cultural Influences as Context}

In the context frame, culture plays a clear role as an example of an instance that can describe the situation and environment of a user. Therefore, a system that includes culture is considered context-aware. Culture applies to a people group. However, country attributes of culture can be considered a reasonable approximation in recommender systems~\cite{Hong2019}.
Researchers have long studied the impact of culture on recommender systems. These studies have applied cultural context to the evaluation and validation of results~\cite{Choi2014}; \cite{Chen2014}; \cite{Chu2017}. This research also suggests that travellers from different cultures and geographies have varying rating behaviours~\cite{Chu2017}. Limited research has included the use of culture as the context in recommender systems~\cite{Huang2021}.
Cultural Heritage was one such study that reviewed how cross-cultural contextualization may be applied within recommender systems. This research utilized rating data, matrix factorization, and clustering techniques to observe how culture affects user preferences~\cite{Hong2019}. This research uncovered that, while often, recommender systems were applied individually, there were cases where aggregate recommendations may also be applied. For example, in cross-cultural contextualization, the research found that country or geographic location could be used as input into the recommender system. 

There have also been a number of studies around the context of user preferences. Hoefstedes model has been used in research outlining musical preferences amongst geographically distributed users and also in user interfaces in mobile recommender systems. They used Hoefstede's model to examine the correlation between similarity in users’ musical preferences. Three specific dimensions of Hoefstede's indices were found to have a notable correlation (masculinity, long-term orientation, and indulgence)~\cite{Liu2018}. In mobile recommender systems studies, uncertainty avoidance and collectivism were most found to influence people’s behaviours and attitudes towards purchase intentions~\cite{Choi2014}.

\subsubsection{Socio-Economic Indicators as Context}
Some recent research in recommender systems has incorporated context in musical systems that add context to traditional content-based approaches by incorporating elements of the user's environment. In these cases, the user's context may include demographic information or information that includes information such as location. In these cases, the location or demographics of the user's area may be paired to make musical recommendations related to select artists from or frequently played in that area~\cite{Kaminskas2012}. In other examples, a particular location's context suggested music typical for that place of interest~\cite{Braunhofer2011}. 
Similarly, demographic indicators were also factored into travel recommender systems, matching user and destination demographic information to provide more accurate hotel recommendations~\cite{Chaudhari2020}. Housing preferences that match users' demographic information for similarities in location have also been an area of research ~\cite{Jun2020}.

\subsubsection{Combining both Socio-Economic and Cultural Context}

Several studies have highlighted the fact that certain tendencies toward traditionally elite entertainment experiences, such as Opera, Theatre, Ballet, Museums, and classical music, have been associated with both various cultural influences and economic status. These are commonly created during the early years of development and are influenced by exposure~\cite{Bourdieu2013}. These studies hypothesize that certain elite economic status enables access to various experiences leading to knowledge and preferences that differ from those associated with lower economic and different cultural backgrounds; this is especially relevant in music~\cite{Kalinowski2021}.

Within music, there have been different studies that have reviewed various contexts surrounding geographic location or the distances between user’s countries, distances between languages of various countries, and socio-economic characteristics such as GDP or Purchase Price Parity~\cite{Schedl2021}, \cite{Liu2018}, \cite{Zangerle2018}. In some cases, these socio-economic dimensions and characteristics have been used to elaborate further on the cultural differences between countries utilizing the World Happiness Report \footnote{https://worldhappiness.report/ed/2022/}. The World Happiness Report utilizes a series of indices that is believed through study to constitute happiness. These indices include gross domestic product (GDP), social support, healthy life expectancy at birth, freedom to make life choices, generosity, perceptions of corruption, positive affect, negative affect, and confidence in national government. In some of these research areas, socio-economic characteristics, in addition to culture, have provided better accuracy in recommender models over those that relied only on learning user preferences based on ratings, selections, and other user-derived attributes. In a number of cases, a feature vector is constructed for each user that includes the socio-economic influences gathered from inputs such as the World Happiness Report, in addition to the scores from Hofstede’s Cultural Indices for the user’s country. This feature vector is then used to find cultural similarities in users’ listening preferences. 

In some studies on music preferences, however, there has been no correlation between purely economic and geographic distances between countries ~\cite{Liu2018}. In these studies, without attributes that represented the cultural differences that influence individual user preferences, recommendations did not improve. This may be because there are varying economic classes within various sub-country geographic levels. The cultural dimensions of masculinity, long-term orientation, indulgence, and the between-country distances in language were found to be positively correlated with an individual's musical preferences.

%%%%%%%%%%%%%%%%%%%%%%%%%%%%%%%%%%%%%%%%%%%
\section{Methodology}

For this research, we define a novel approach to understanding the socio-economic-cultural distances between countries that may be relevant in streaming media entertainment. We use a combination of supervised, and unsupervised machine learning techniques, an ontology that includes a number of socio-economic-cultural indicators, as well as economic and opportunity indicators that are specific to streaming video. We determine the relevant features that have the highest impact and use these features to create various representative vectors that can be incorporated in order to provide a context within a recommender system. In our motivating scenario, we review the application of how a user in a specific country may have a certain preference in a subscription plan for streaming based on other users that are in a similar socio-economic-cultural country. We use a combination of key indicators to calculate this specific similarity in streaming media between countries.

\begin{figure} [!ht]

    \centering

    \includegraphics[width=1\textwidth, angle=0]{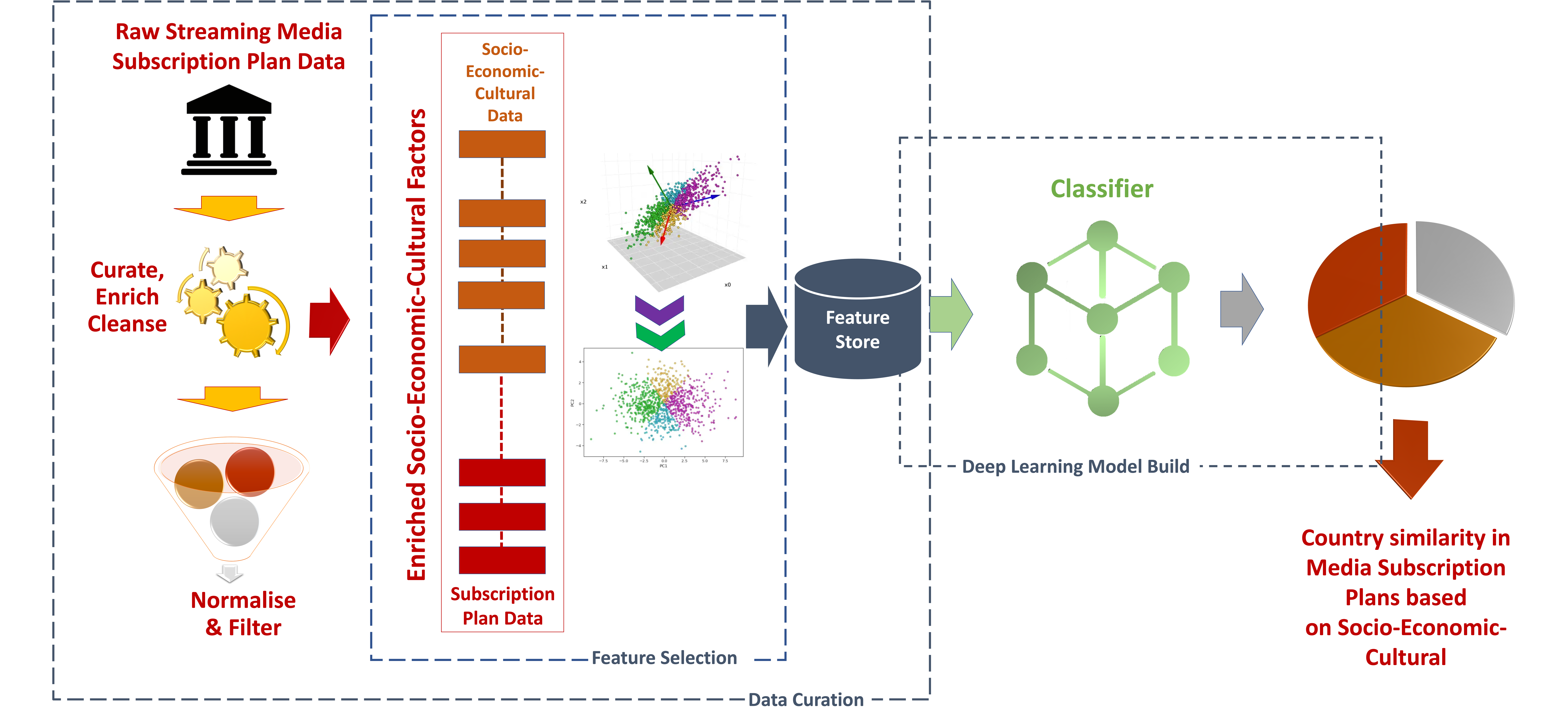}

    \caption{Data, Feature and ML Process for Understanding Socio-Economic-Cultural factors for Streaming Video Subcriptions}

    \label{fig:Data Processing}

\end{figure}

We then utilize a series of steps to pre-process our data, curate and enrich our data, normalize and filter our data, reduce our data to key features, build and train a model to classify data, and then evaluate our model. These will be discussed in more detail but a high-level visual of this process can be seen in Figure:~\ref{fig:Data Processing}. Data in this process is extracted, curated, normalized, and filtered. We then enrich the domain data with socio-economic and cultural data, select and store our features, and then model them into a final classifier. The final feature set is what we are able to generalize into recommender systems for media-based subscriptions. 

\subsection{Data and Data Processing}

\textbf{Data Pre-processing}.
We use subscriber data that represents the plans and pricing selection of users within a country for a streaming media provider as our domain data. As pricing becomes most relevant when users have displayed a self-elected change in plan and price over a period, or subscribe and unsubscribe to a service over a period. Therefore, we focus on cohorts who have self-elected a change, have left, and then resubscribed to the service at least once. Users who have left and re-subscribed to the service at least once, represent a group of users where potentially the rising costs of subscription services relative to perceived value in an increasingly competitive environment is a potential reason for the user to leave the service vs maintain a service~\cite{Kweon2021}.

We narrow our cohort to understand and try to predict the change in plan and price after the media service has invoked a price increase. A single user may leave, return to the service (re-subscribe), change plans, or incur a price change. Our pre-processing of the data includes aggregation by a user’s account, country, number of plan changes, and the resulting from and to plan change. We used this data to create a base matrix representing each country's features. In our scenario, this final matrix will include attributes from our domain data, such as 1) the population for each plan, 2) corresponding pricing, 3) the data of the most recent plan selection, and 4) the corresponding country. As our research is at the country level, we eliminate the need and use for any personally identifiable data in our dataset, by constructing the data at an aggregate level. 

\textbf{Curate, Enrich, Cleanse}. 
In our domain data, we have all the data for each country. This minimizes some amount of cleansing. However, depending on when new plans are introduced within a country, the timing of various marketing strategies that drive particular plan selections, or changes in business strategy, our data may not be consistent. To ensure a strong data set, we review the data and calculate various statistics around our data, such as variance and mean. These statistics assist to identify and enable the removal of outliers. 

For each country represented in the domain data, we enrich that country’s data with data made available through various publicly available knowledge bases. The enriched streaming video plan data describes the country with additional cultural, economic, demographic, geographic, infrastructure, and media-related data attributes for each specific country. The potential feature data, from various knowledge bases, is not available for all countries. This means that not every feature will be available for each country. In this case, we use techniques that include removing data, predicting potential substitutes, resorting to the last known observation (in our case the previous year’s data), or imputing data with the mean using machine learning or other statistical methods.  

\begin{figure}[!ht]   %[ht!]
    \centering
    \includegraphics[scale=1.8, angle=0]{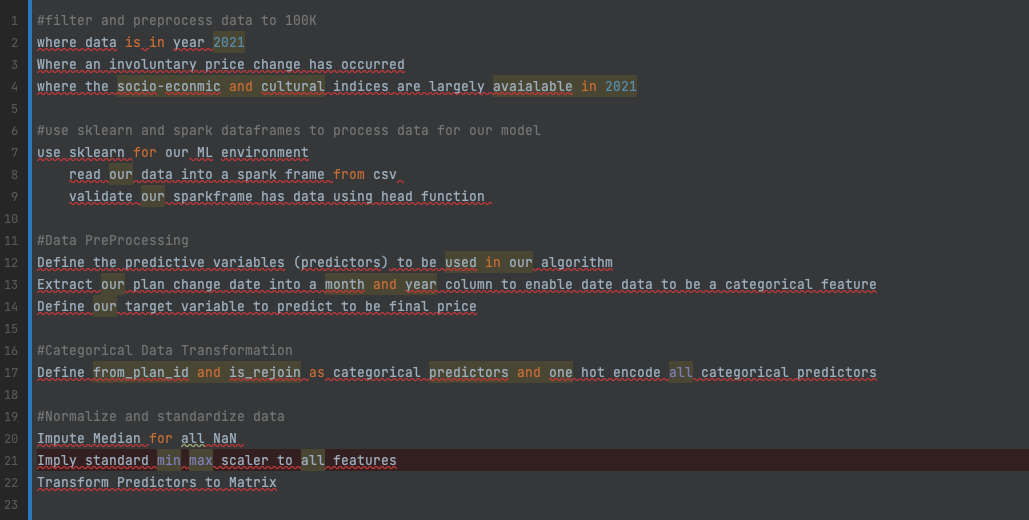} %width=1\textwidth,
    \caption{Data Preprocessing Pseudocode}
    \label{fig: Data Preprocessing}
\end{figure}

\textbf{Normalize and Filter}.
To minimize variance in constructing our vector, we normalize our features by applying a logarithmic function that normalizes our data between zero and one. In some cases, we apply a filter to adjust the range selection of our data. This filtering may be based on domain-related knowledge or to select data most relevant to our model. For example, within the economic factors, the data may have existed for multiple years in some of our public domain data. However, only the most recent data provides predictability in the current year. 
As we classify our data using Neural Networks, features with similar scales help converge gradient descent to a local minimum significantly quicker than those that do not~\cite{Djordjevic2022}. To do this, we apply a minimum and maximum scaler function. 
We can see a sample of the preprocessing steps in the pseudocode below \ref{fig: Data Preprocessing}.

%\vspace{0.1cm}
%\vspace{0.1cm}

\subsection{Feature Selection}

We can group the variables identified with potential impact as cultural factors, economic factors, demographic factors, and geographic factors. We include country-specific infrastructure data related to broadband internet, mobile networks, mobile and tv devices, and other media-related data attributes for each specific country. These factors provide a potential indication of a particular country’s readiness and opportunity for streaming video. 

Since our attributes are numerical, we skip feature extraction and move to reduce our feature set through the selection of only highly relevant features. To understand which features are highly relevant, we use apply methods to select features that may have the best impact on predictability. To achieve this, we organize our input data by describing the data within an ontology. Our ontology describes a hierarchical class structure grouping related attributes. 

\subsubsection{Feature Reduction through Correlation}

\begin{figure}[!ht]   %[ht!]
    \centering
    \includegraphics[scale=1.8, angle=0]{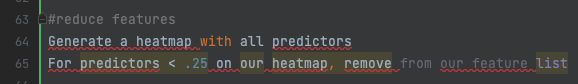} %width=1\textwidth,
    \caption{Feature Reduction Pseudocode}
    \label{fig: Feature Reduction Pseudocode}
\end{figure}

Heatmaps are visualization tools that can allow gradients of colour across both the Y and X axis to visually display relationships between contexts. We use an approach similar to Buyrukoğlu~\cite{Buyrukoglu2022} and use a heat map to determine correlations between features and specifically our target prediction with our classifier. This approach allows us to examine the relationships between each feature and our target prediction, enabling us to reduce our feature selection to those with the highest perceived impacts. We reduce our features sets by removing features with less than .25 correlation (Figure: \ref{fig: Feature Reduction Pseudocode}).

%\vspace{0.1cm}
%\vspace{0.1cm}

\subsection{Modelling Neural Network for Price Prediction}

In this chapter, we reviewed the effectiveness of socio-economic-cultural aware features with domain data to predict consumer preferences in price. This allowed us to understand the potential positive impact of use in recommender systems. We found these features to positively impact price prediction with relative accuracy. 

\begin{figure}[!ht]   %[ht!]
    \centering
    \includegraphics[scale=1.0, angle=0]{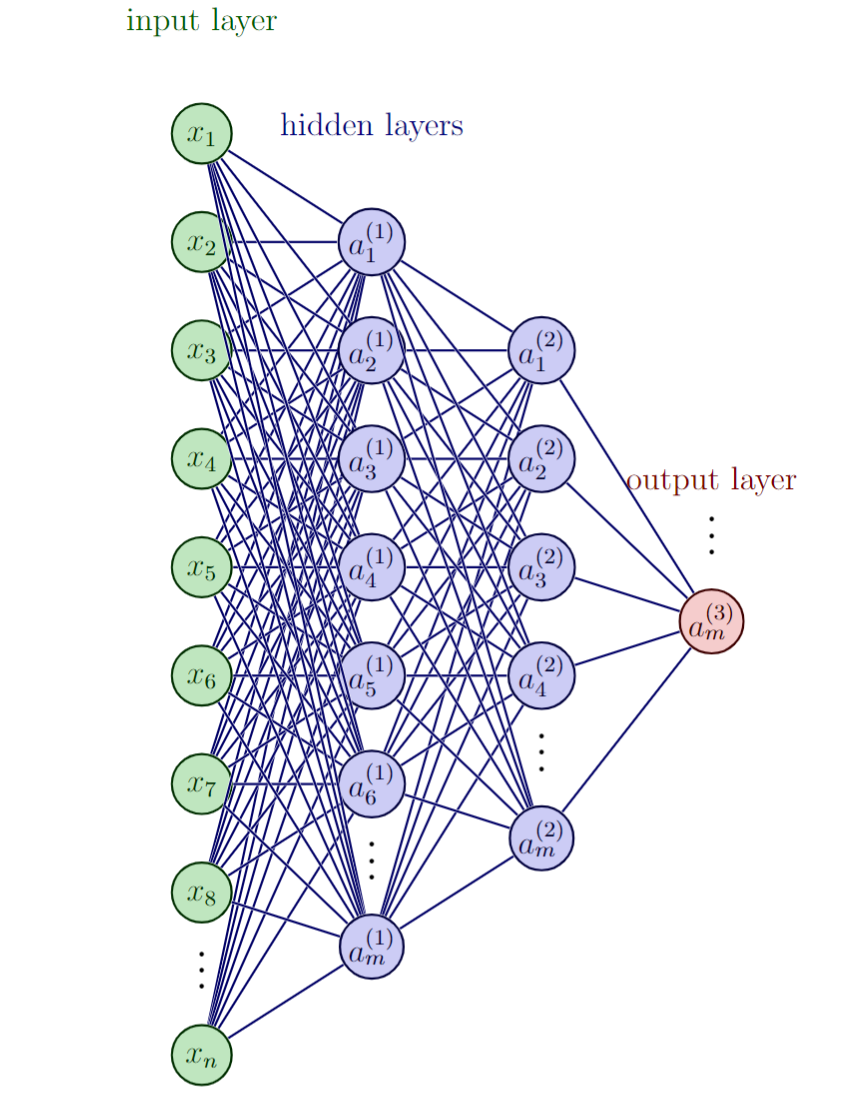} %width=1\textwidth,
    \caption{Artificial Neural Network with two hidden 
 layers, and a single output}
    \label{fig: Artificial Neural Network}
\end{figure}

We also found that the most effective machine learning model in prediction was a well-architected artificial neural network model with \textit{select features}. We also learned that, with a well-architected neural network model, the feature set could be reduced, providing computational efficiency with similar accuracy.  For the artificial neural network model, we chose a neural network model with one input layer, two hidden layers, and a final output layer, with a single neuron. Figure ~\ref{fig: Artificial Neural Network} references an example neural network with two hidden layers, each layer reducing the number of inputs and resulting in a final output layer.

To find the optimal batch size and epoch size, we run a process that tests the permutations of different batch sizes and epoch sizes. We use mini-batch sizes at multiples of two, and epoch sizes starting at a maximum of three times the number of predictive features in our input layer and then reducing. 
We use adaptive moment estimation or \textit{’Adam’} as our optimization algorithm in our neural network to update our weights iteratively. Adam adapts the learning rate with each training batch. This approach allows us to focus on optimizing our cost function without concentrating on selecting the optimal learning rate.

We use the following three rules of thumb \cite{Heaton2015}to determine the number of neurons in each of our hidden layers:  

\color{black}

\begin{itemize}

    \item the number of neurons in a single hidden layer is somewhere between the number of features used in the input layer, or previously hidden layer and the total number of predicted outputs in the output layer

    \item the number of neurons is the total number of predicted outputs in the output layer added to approximately two-thirds the size of the previous input layer

    \item the number of neurons is less than twice the size of either the features used for input or the previous hidden layer

\end{itemize}

\begin{figure}[!ht]   %[ht!]
    \centering
    \includegraphics[scale=1.8, angle=0]{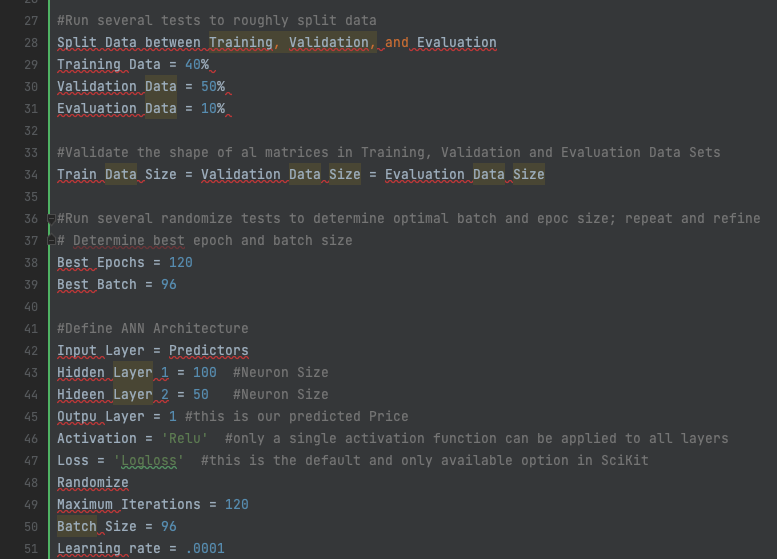} %width=1\textwidth,
    \caption{Initial ANN Pseudocode}
    \label{fig: ANN Pseudocode}
\end{figure}

We split our data for training, validation and evaluation. Then, we establish our initial ANN architecture parameters, inclusive of two hidden layers, the number of neurons in each layer, the batch and epoch size, the loss function, optimizer, learning rate, and activation function (Figure: \ref{fig: ANN Pseudocode}).

%\vspace{0.1cm}
%\vspace{0.1cm}

%https://www.ncbi.nlm.nih.gov/pmc/articles/PMC8459779/

%https://iopscience.iop.org/book/mono/978-0-7503-2216-4/chapter/bk978-0-7503-2216-4ch3

\subsection{Optimizing our Model Architecture}

\subsubsection{Preventing Over-fitting}

In machine learning, when our model’s prediction may be inaccurate, one reason this occurs may be based on how the training data fits the model. If too many features are incorporated into the model, it may memorize the values of the training data and fail to generalize the data when applying the model over time to new examples. We know this state of memorization as over-fitting~\cite{Ying2019}. Similarly, when our model is too simple for our data, it may bias toward certain training data values, also failing to generalize the model to fit different examples. This usually occurs when there are insufficient features or data to train the model. This is called underfitting~\cite{Jabbar2014}.

\begin{figure}[!ht]   %[ht!]
    \centering
    \includegraphics[scale=1.0, angle=0]{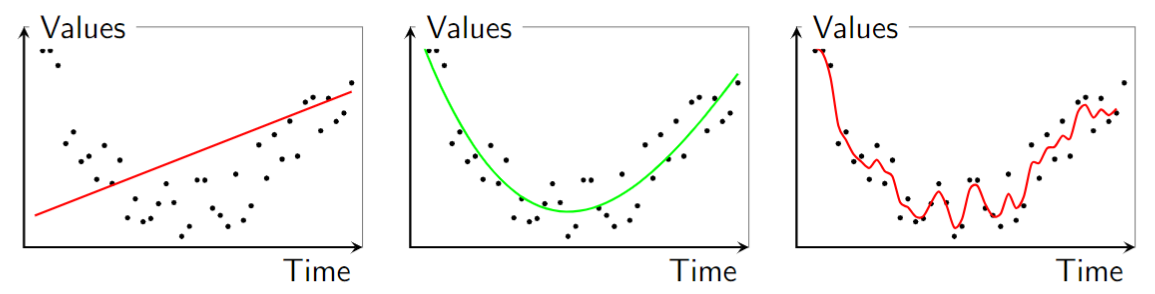} %width=1\textwidth,
    \caption{Fitting a line to our model (Underfitting, Right Fitting, and Overfitting)}
    \label{fig:Model_Fit}
\end{figure}

In our research, we aim to have enough data to reduce the likelihood of underfitting by increasing the data volume. Overfitting our data, however, remains a risk. To ensure that we do not over-fit our model, we use \textit{dropout}~\cite{Srivastava2014} as a mechanism to regulate our models in each layer. This regularization technique drops a specified number of neurons with each training iteration. Dropout techniques work only for robust numbers of features. Each iteration is equivalent to training a separate model by dropping different neurons in each iteration. In this case, envisioning each iteration as a different network means each network will over-fit in different ways. Since each network over-fits differently, the aggregate of all the different iterations reduces over-fitting.

Figure~\ref{fig:Model_Fit} displays the three categories of fit, where in the first example the model underfits the variance of found in data over time, a second example where a model fits data over time reasonably well with little variance, and the final example where a model becomes hyperaware to every variation in the data leading to potential issues and inefficiencies over time.

\subsubsection{Optimization and Activation}

Neural networks train at different learning rates depending on the algorithm. This learning rate is the rate at which our algorithm learns and adapts to new training data. In machine learning, as our data learns, it converges to a local minimum that represents the best possible accuracy in our prediction. For our model, we choose the Adam optimizer as an optimizer that adapts the fastest to learning our training data. Adam optimizers are computationally efficient and adapt effectively for large feature sets or training data sets~\cite{Kingma2017,Beheshti2018a}.

In neural networks, the neurons and layers represent weights and biases. In each layer of a neural network, and with each neuron, these weights and biases are carried into the next stage as input. The technique is similar to those within the human brain and its methods of learning. Activation functions represent the method by which each of these weights and biases may be calculated~\cite{Jones2014}. 
We test several activation functions, including using a single activation function for all layers and, eventually, using tailored and different activation functions for each layer in our model. In our final model, we explore the use of the \textit{Sigmoid} function. Our output is a price prediction. The \textit{Sigmoid} activation function converges to a single output, making its use in our model logical to test.

The \textit{Sigmoid} activation function works by reducing our inputs \textbf{(x)} to a range between 0 and 1. It is typically used in binary classification, or regression when a model output is a certain prediction~\cite{Nwankpa2018}. 

\begin{equation}
   f(x) = 1/(1 + exp(-x))
\end{equation}
\label{eq:3.1}

3.1 Sigmoid Activation Function.
Sigmoid Activation Function

We explore the use of a \textit{Relu}, Rectified Linear Unit, activation function in one of our hidden layers. \textit{Relu} activations are well-adapted and perform optimally for neural network hidden layers \cite{K.He2015}. 

\begin{equation}
    f(x) = max(0,x)
\end{equation}
\label{eq:3.2}

3.2 Relu Activation Function.
Relu Activation Function

And lastly, we experiment with the \textit{Tanh} activation function, \textit{Tanh} activation functions follow a more compressed curve than a \textit{sigmoid} activation function and therefore are known to perform more efficiently with multilayer models~\cite{Karlik2011}. They are often used in speech-to-text neural networks. 

\begin{equation}
    f(x) = tanh(x)
\end{equation}
\label{eq:3.3}

3.3 Tanh Activation Function.
Tanh Activation Function

\subsubsection{Evaluation of our Model}
We evaluate our classification model by reviewing metrics such as Accuracy, Precision, and F1. These metrics measure the impacts of what is otherwise known as a confusion matrix~\cite{Liang2022} and the different elements within the matrix relative to the others. In a confusion matrix, predictions are described as \textbf{True Positives (TP), True Negatives (TN), False Positives (FP), and False Negatives (FN)}. Measures utilizing these descriptions are used to evaluate how well a model performed. 

%\vspace{1cm}
Accuracy represents the efficacy of how accurate our model predictions were by comparing true positives and true negatives to the total number of predictions made. 

\begin{equation}
 Accuracy = \frac{(TP+TN)}{ (TP+TN+FP+FN)}  
\end{equation}
\label{eq:3.4} 

3.4 Accuracy. Accuracy

Precision represents how often our positive predictions were correctly made compared to all our positive predictions. 

\begin{equation}
Precision = \frac{TP}{ (TP+FP)} 
\end{equation}
\label{eq:3.5} 

3.5 Precision. 
Precision

The recall represents the efficiency of our predictions by comparing our true positive predictions to all the predictions that should have been positive.

\begin{equation}
 Recall = \frac{TP}{(TP +FN)}
\end{equation}
\label{eq:3.6} 

3.6 Recall. Recall

F1 takes into account both precision and recall, noting the trade-off between the two.  The equation can be described as mathematically as either: 

\begin{equation}
F1 Score = \frac{TP}{TP +\frac{1}{2}(FP+FN)} 
\end{equation}
\label{eq:3.7} 

F1 Score

\begin{center} or \end{center}
\begin{equation}
F1 Score = 2* \frac{Precision * Recall}{Precision + Recall}   
\end{equation}
\label{eq:3.8} 

%Alt F1 Score.
%\vspace{1cm}
3.7 \& 3.8 F1 Score

We also compare our model to other classification machine learning models such as Stochastic Gradient Descent, Random Forest, and Gaussian Naive Bayes.

%%%%%%%%%%%%%%%%%%%%%%%%%%%%%%%%%%%%%%%%
\section{Experiment, Results and Evaluation}

\subsection{Motivating Scenario: Streaming Media Domain}

Our project aims to understand the relationship between streaming media plan selection and cultural and socio-economic indicators. We use an aggregation of customers across a few selected countries. These particular customers have changed subscription plans after a price increase has occurred. These selected countries are based on available cultural and socio-economic indicators. To represent the reaction to a plan change, we isolate customers who instantiated what would be considered a down-graded plan change within 3mos of the price increase. We aggregate customers within a country to understand the frequency of plan changes, and the similarities between the types of plans that these users were on and downgraded to. 
Our model uses three classes of features, all aggregated at the country level. These three classes of features are streaming media domain features, cultural indicators, and socio-economic features. 

%\vspace{0.1cm}
%\begin{figure*}[!ht]   %[ht!]
%    \centering
%    \includegraphics[scale=2.8, angle=0]{ch_exp/figures/Final Features before Encoding.png} %width=1\textwidth,
 %   \caption{Final Set of Domain and non-Domain Features}
%    \label{fig: Final Features}
%\end{figure*}

\subsubsection{Domain Features}
For domain-related features, we extract customer categorical and non-categorical numerical features available from a media-related company. This data is aggregated by country, the type of plan change, the date of sign-up, and the date of the plan downgrade. We review only downgraded plan data in a single year. We use the domain’s experts’ and streaming media managers’ points of view about each feature. The main features used in our scenario are:

\begin{itemize}
    \item The number of accounts at a country level that has downgraded plans %on a specific day and within a specified time period
    \item The total number of times a customer changed plans throughout the customer history
    \item Whether they have ever left, they abandoned their subscription and returned to the service at least once
    \item The type of plan change that occurred, and what the to/from plan type and price were at the time of change
    \item The number of customers in a given country that enacted a plan downgrade on the same specified day   
\end{itemize}

\subsubsection{Cultural Features}

We use Hoefstede's cultural model and indices, as noted earlier in our paper. Hoefstede's model provides country-level features that enable the comparison of countries across range-mapped cultural indices, such as 
Long-Term Orientation (LTOWVS),
Power Distance (PDI),
Individualism (IDV),
Masculinity (MAS),
Uncertainty Avoidance (UAI), and
Indulgence (IVR).

%\begin{itemize}
%    \item Long Term Orientation (LTOWVS)
%    \item Power Distance (PDI)
%    \item Individualism (IDV)
%    \item Masculinity (MAS)
%   \item Uncertainty Avoidance (UAI)
%    \item Indulgence (IVR)
%\end{itemize}

These indices provide spatial distance between countries to help understand cultural differences. For instance, Japan has a score of 88 for long-term orientation, compared to Columbia, which has a score of 13. This represents a more significant societal value placed on decisions and values that have an impact across long periods for most people in Japan, as compared to Columbia, whose score indicates a more short-term orientation in values and decisions. We aim to understand how the cultural indices at a country level may impact any of the media subscription attributes for customers within that country. 

\subsubsection{Socio Economic Features}
For our social and economic features, we gather publicly available information across several areas. These areas are both domain and non-domain-relevant. Relevant indicators are typical in understanding a given country’s demographic, relevant infrastructure, economic status, or economic investments. Entertainment indicators are used to describe our domain data. We concatenate this data with our country-specific level data both at the country level and the streaming media domain level. We aim to find the relationship between a country’s socio-economic data and customer actions in streaming media.  
We group this data across four main categories for each country:

\begin{itemize}
    \item \textbf{Infrastructure} - the general network communication and broadband data available within that country.  This data consists of broadband subscriptions, mobile network coverage, and communication infrastructure investments.
    \item \textbf{Demographics} - which includes data related to population, and graduates in stem
    \item  \textbf{Economic Indicators} - indicators of a country's economy such as gross domestic product, foreign exchange rate, and spending on communication infrastructure
    \item \textbf{Market Opportunity} - These features represent potential relevance to the opportunity for streaming media within a given country. They are indicators such as the population streaming television, the average screen viewing hours, global brand value, national films produced, and spending in entertainment.  
\end{itemize}

%\vspace{0.1cm}
\begin{figure}[!ht]   %[ht!]
    \centering
    \includegraphics[scale=2.5, angle=0]{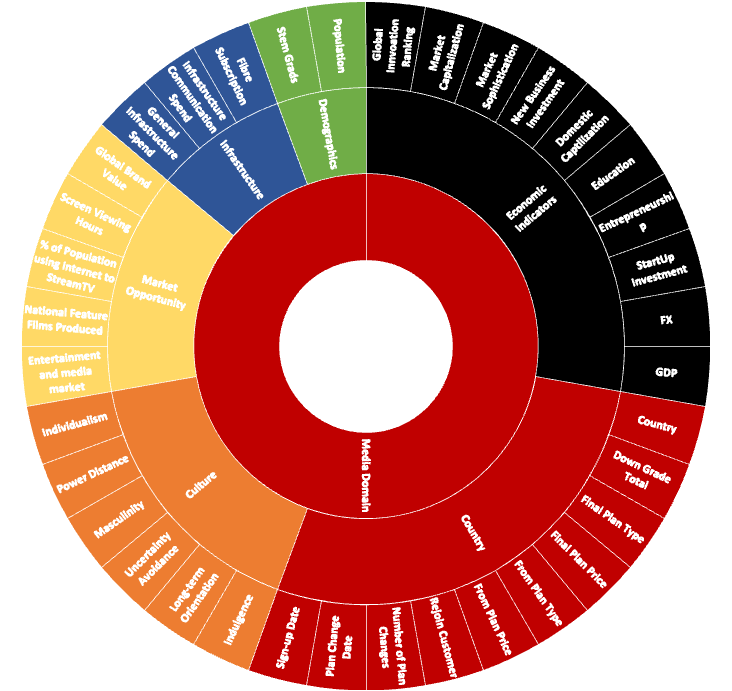} %width=1\textwidth,
    \caption{Domain and Non-Domain Specific Features}
    \label{fig: Features}
\end{figure}
%\vspace{0.1cm}

\subsection{Dataset}
For our data set, Figure~\ref{fig: Features}, we compile a list of open data from various sources with globally relevant socio-economic data on all or most countries. These data sets included sources such as those from World Bank \footnote{https://www.worldbank.org/en/home}, UN level data \footnote{http://data.un.org/default.aspx}, Organization for Economic Co-operation and Development \footnote{https://www.oecd.org/}, and Statista \footnote{https://www.statista.com/}. This socio-economic data includes 36 features and up to 195 unique labels, representing the number of potential countries. We use publicly available data from Hoefstedes’ cultural indices \footnote{https://geerthofstede.com/research-and-vsm/dimension-data-matrix/} across all available countries. The cultural data represents the full six indices and 112 countries. The data was last updated in late 2015. For our domain data, we use a sampling of data across 14 specific countries scattered globally across various continents. The domain data in this sample set represents approximately 100,000 rows and 7 features.

%\vspace{0.1cm}
\begin{figure}[htpb!]   %[ht!]
    \centering
    \includegraphics[scale=1.0, angle=0]{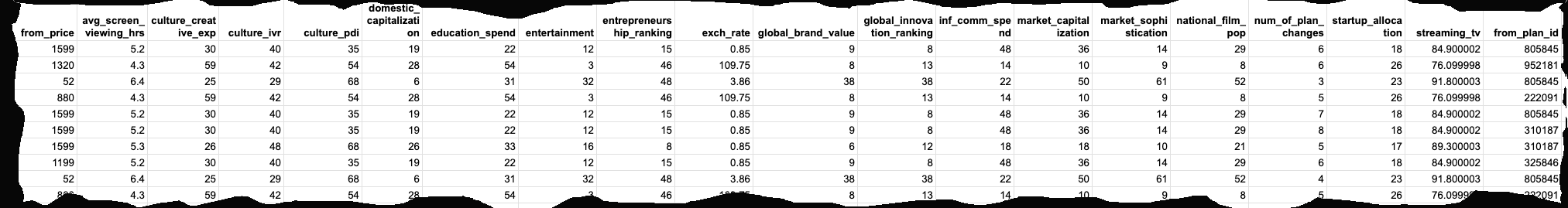} %width=1\textwidth,
    \caption{Sample Data from a selection of the extracted features}
    \label{fig: Data}
\end{figure}
%\vspace{0.1cm}

\subsection{Pre-Processing}
As the data we retrieved is public, the quality and consistency of the data need to be evaluated. Data also is pre-processed in order to ensure the highest quality of data inputted into our Machine Learning models. This pre-processing enables the highest accuracy in our predictions.  

\subsubsection{Removing and Filtering Inconsistent Data}
\begin{figure}[!ht]   %[ht!]
    \centering
    \includegraphics[scale=1.18, angle=0]{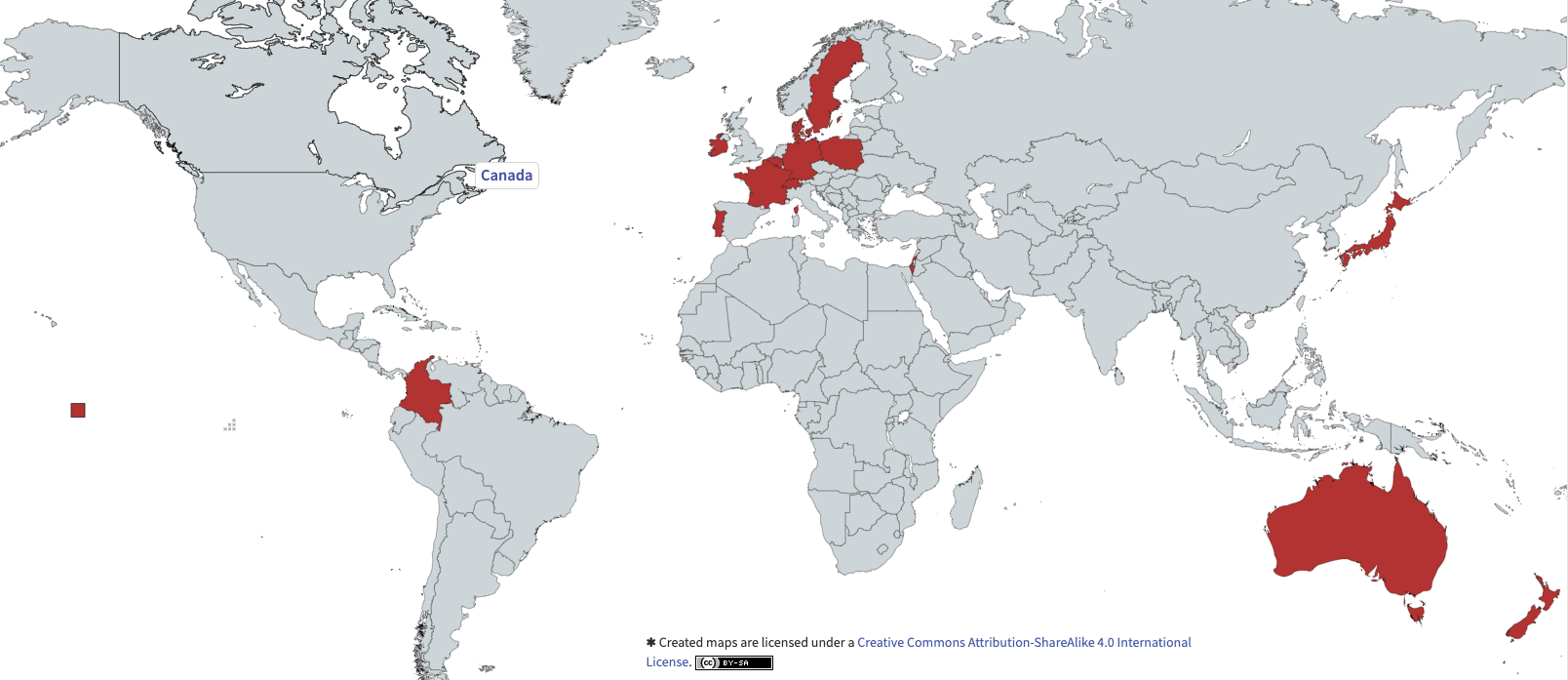} %width=1\textwidth,
    \caption{Our 14 Countries with Full Cultural and Socio-Economic Features.}
    \label{fig:Countries}
\end{figure}

We denote numerous inconsistencies across data our socio-economic data. In particular, a wide range of absent or stale data exists for certain indicators across countries. Rather than approximating this data with techniques such as mean, min, max, or zero, we utilize this opportunity to reduce our country feature set, and more densely orient our domain data to a few countries. We filter only recent data and join all of our publicly available data with an inner join. This reduces our feature set to 14 specific countries (Figure: \ref{fig:Countries}) and drops all countries with absent or stale data. 

%\vspace{0.1cm}
%\vspace{0.1cm}

We also remove features in our domain data set that may be too predictive of our final result but will not generalize well. For instance, data such as total downgraded subscribers for a specific date, and country, only indicates to us that there was a collective reason on that day to downgrade their plan. As we prefer not to have our model be hyper-aware of temporal differences at this stage, we remove this. In future work, more specific features that map to temporal differences may yield better, and more specific, predictions. Our current non-domain data set is limited in its awareness of temporal factors, and as a result, remove this data.  

\subsubsection{Standardizing Numerical Data}

As the range of data attributes across our data varies, and in some cases, the mean or variances of specific features may differ significantly from one country to another, we standardize our data to minimize any potential deviations that features may have either with each other or from country to country. 

\subsubsection{Normalizing Data}

We normalize our data or otherwise potential features to be between 0 and 1 to ensure that we fit our model well and do not inadvertently create biases with large variations in our data. We do this by using a linear transformation of our input value x. In particular, we perform min-max scaling \cite{Dey2018}, where we take the minimum feature value \textbf{min(x)} found and the maximum feature value \textbf{max(x)} found in our data and utilize the differences and the difference with our input value, to arrive at a representation of our original input value x. This new representation of x is consistently between 0 and 1.

\begin{equation}
 x^1 = ~^x scaled = \frac{(x-min(x))}{ (max(x)-min(x))}  
\end{equation}
\label{eq:3.9} 

Normalizing Data: Min-Max Scaler

\subsubsection{Categorical Data Transformation}

For our domain-based data, we identify some data as categorical. Categorical data identified is data such as which plan the user subscribes to, and the information used to determine whether a customer has left and resubscribed to the service. Our domain data also contains key date information. For date information, we separate out month and year. We make use of the technique of one-hot encoding to classify our data into binary columns. We add these new binary columns to our feature set and drop our previous ones.

%\vspace{0.1cm}
\begin{figure}[!h]   %[ht!]
    \centering
    \includegraphics[scale=1.7, angle=0]{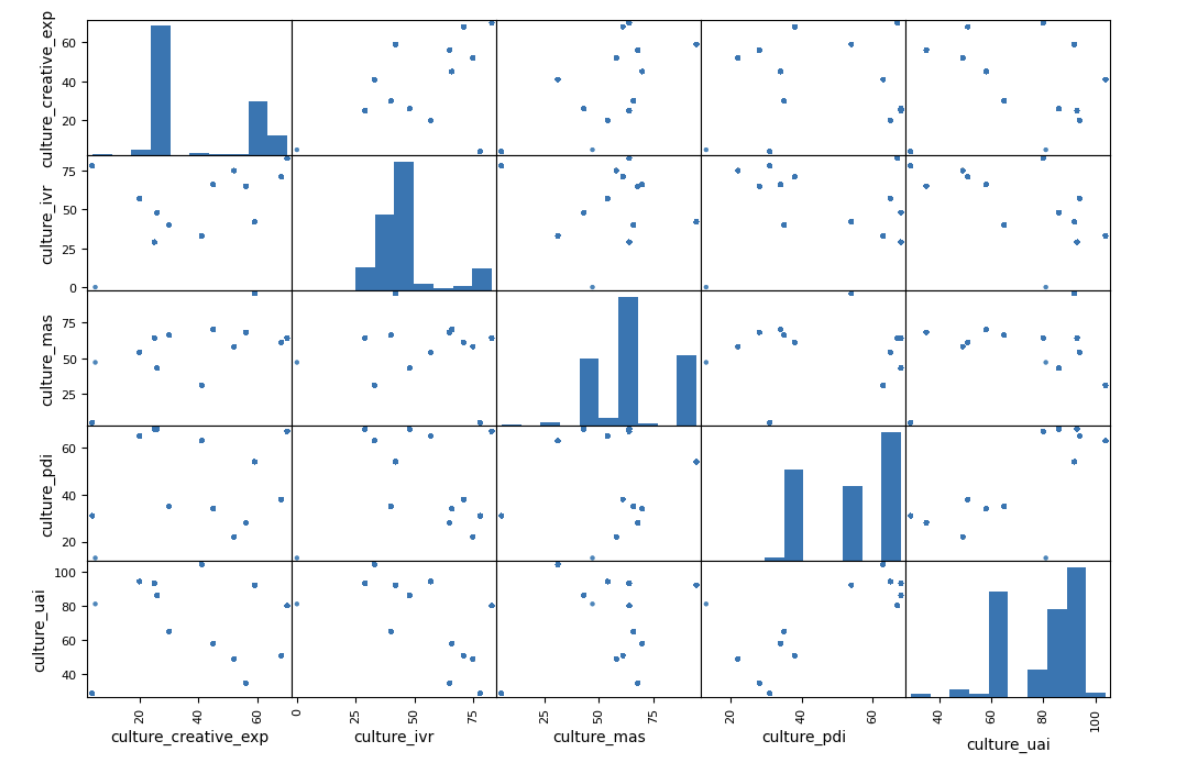} %width=1\textwidth,
    \caption{Displaying culture correlations using a scatter matrix across countries}
    \label{fig:Evaluating Culture Values across Countries}
\end{figure}

%\vspace{0.1cm}
\subsubsection{Evaluating diversity of our Cultural Data}

Since culture is a key component of our work, we evaluate the distribution of diversity across our 14 countries and their indices. This shows that our cultural data has little overlap and is appropriately spread across various indices. We use a scatter plot to visualize the distribution. With the scatter plot we construct, we are able to understand that our country choices vary well for our modeling exercise.

\subsubsection{Splitting our Data}

For our model, we split our data into several test data sets. We randomize the data to ensure that it has the possibility of best representing the full data set as much as possible. In our final neural network, our data is divided up into three parts: training data, test data, and validation data. We use our validation data to determine the hyperparameters in our neural network. We use a 30\% split of our data across each data set. 

\subsection{Feature Reduction}
\begin{figure}[!ht]
  \begin{subfigure}[b]{0.52 \textwidth}
    \includegraphics[width=\textwidth]{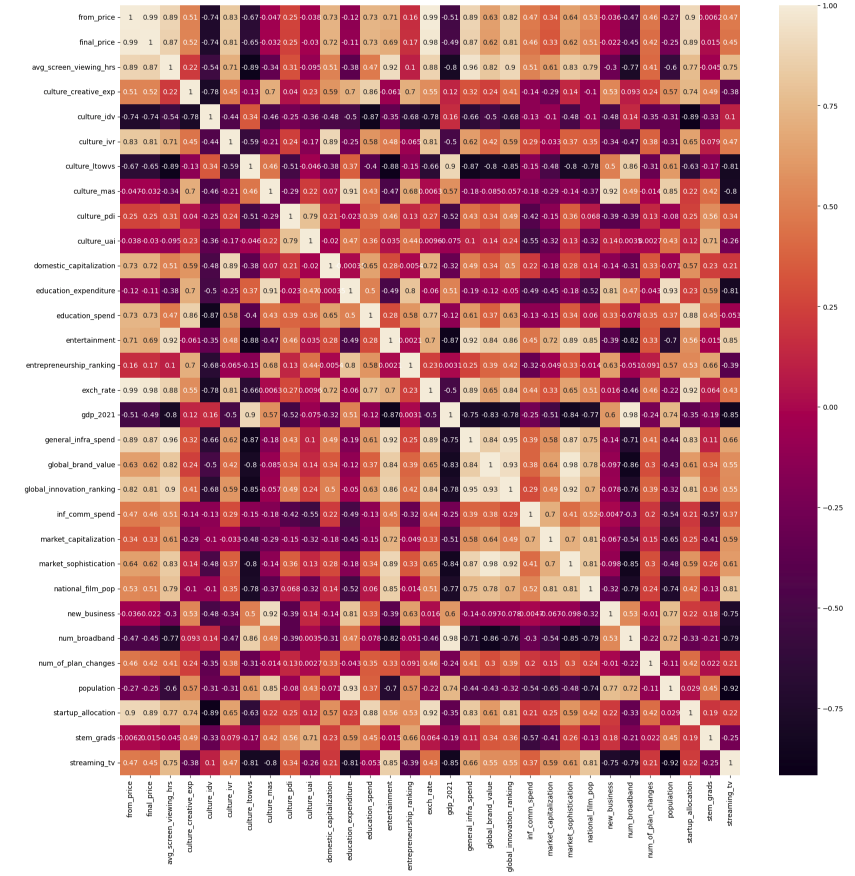}
    \caption{Original Feature Set}
    \label{fig:Original Feature Set}
  \end{subfigure}
  \hfill
  \begin{subfigure}[b]{0.5 \textwidth}
    \includegraphics[width=\textwidth]{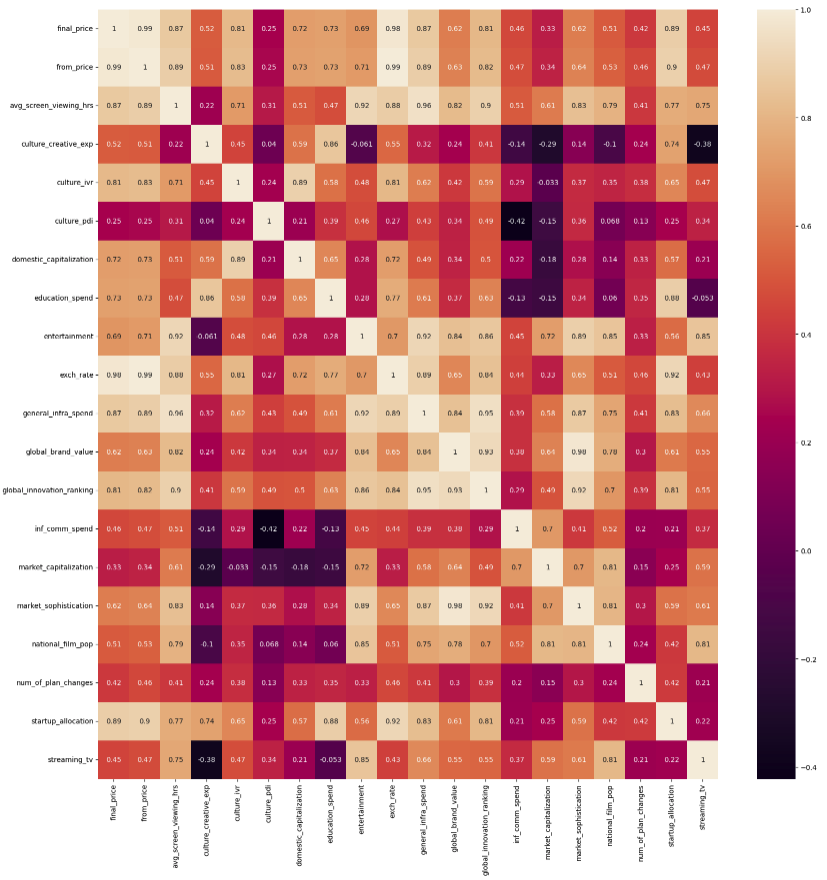}
    \caption{Final Feature Set}
    \label{fig:Final Feature Set}
  \end{subfigure}
\vspace{-0.2cm}
  \caption{Correlation Heatmap for Feature Reduction}
  \label{fig: Correlation Heatmap}
\end{figure}

We use price as an indicator of preferred plan selection in our population. As this is our predicted output in our model, we craft a correlation heatmap to evaluate our feature’s relationship with our predicted output. We use this heatmap to understand which primary features that overlap in our data set. We eliminate any feature whose correlation with our final price is a scored value less than .25.  Scores less than .25 are indicative of low predictability in our model. We also eliminate values that seem to correlate highly with other features but with less feature importance to our final output. Our final feature set can be reviewed in the heatmap shown in Figure~\ref{fig: Correlation Heatmap}.% or in Appendix: Chapter~\ref{chap:appendix}.

%\vspace{-0.1cm}
%\vspace{-0.1cm}

After feature reduction, our final selected domain and non-domain features are represented in Table
~\ref{tab: Final Features} with a brief description of each attribute.

%\vspace{0.1cm}

\begin{table}[h!]
        \centering
        \includegraphics[scale=0.38]{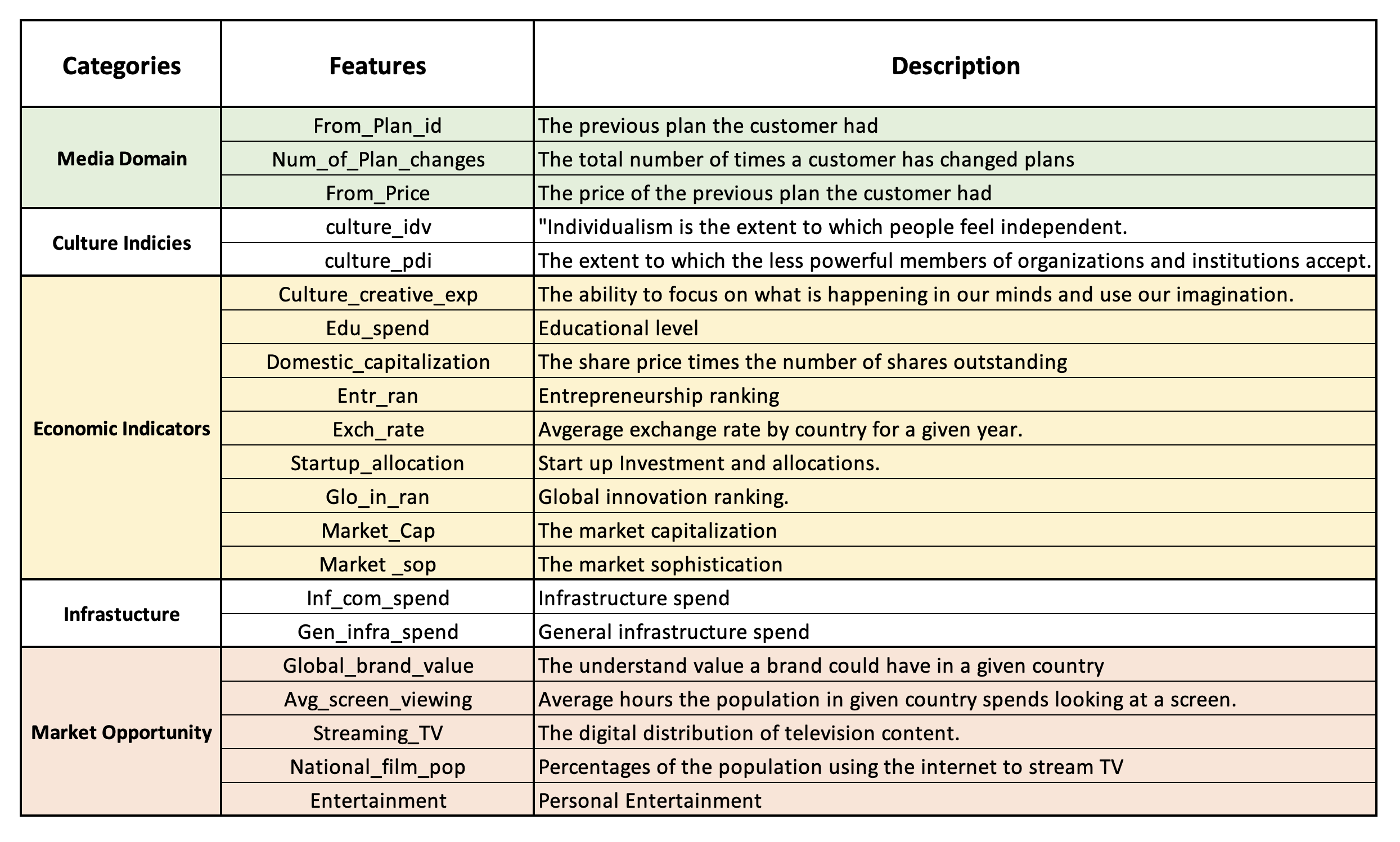}
        \caption{Domain and non-domain based final selected features after feature reduction }
        \label{tab: Final Features}
\end{table}

\subsection{Classification}

With our data cleansed, normalized, standardized, and all our features selected we use this data as the input to our model.  With one hot encoding our features after feature reduction total 71 inputs to our model.  We utilize \textit{Scikit-learn}\footnote{https://www.openhub.net/p/scikit-learn}, an open-source machine learning package, and \textit{Python} to test our classification

\subsubsection{Neural Network Model}
For our first artificial neural network model, we use a fully connected artificial neural network with two hidden layers. In our first hidden layer, using the third rule of thumb where are neurons are less than twice the size of our features for input. We utilize 100 nodes in our first hidden layer. For our second hidden layer, we use our first rule of thumb and select a number between our first hidden layer and the number of outputs in our output layer. Since we are reviewing price as an indication of the plan, we have only one output in our model. Our second hidden layer contains 50 nodes.

\textbf{Activation and Loss Functions}.
We utilize \textit{ReLu}, or rectified linear activation function, as our activation function for our first ANN model. This linear function outputs a zero if the result is negative, else if the result is positive it passes the input it received as the input into the next layer.  We also test a logistic sigmoid function.

\begin{figure}[ht!]
    \centering
    \includegraphics[width=1\textwidth, angle=0]{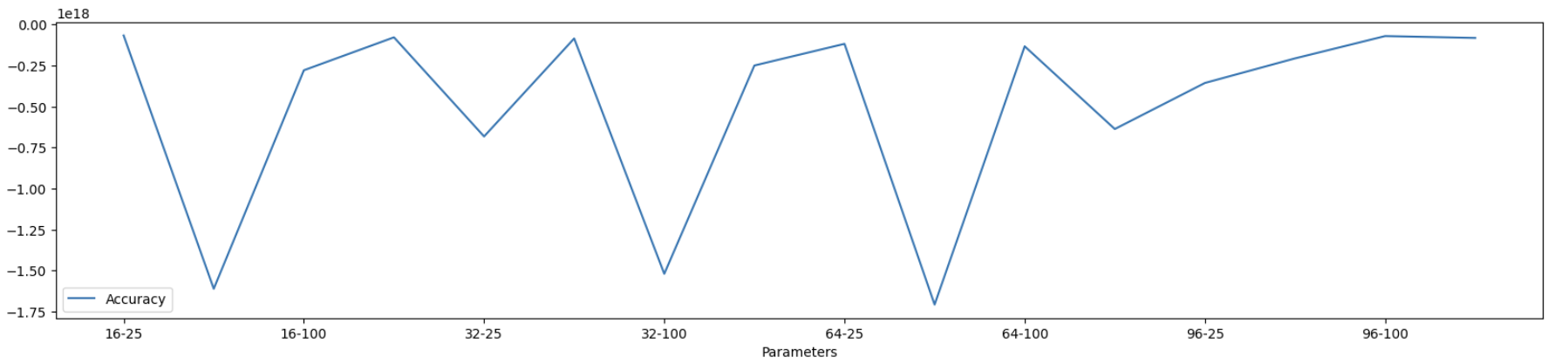}
    \caption{Evaluating Hyper Parameters}
    \label{fig:Evaluating Hyper Parameters}
\end{figure}

\textbf{Evaluation of Hyper Parameters}.
To understand the right hyperparameters, epochs, and batch sizes in particular we should use to train our model, we test several permutations. We use batch sizes of 16, 32, 64, and 96, and epoch sizes of 25, 50, 100, 120. We graph our output (Figure: ~\ref{fig:Evaluating Hyper Parameters}) and see that our model performs best near a batch size of 96 and epochs between 100-120. Based on this we select a batch size of 96 and epochs of 120 to train our model. 

%\vspace{-0.1cm}

%\vspace{-0.1cm}

\subsubsection{Evaluating other Machine Learning Models for Classification}
\begin{figure}[!h]
  \begin{subfigure}[b]{0.5\textwidth}
    \includegraphics[width=\textwidth]{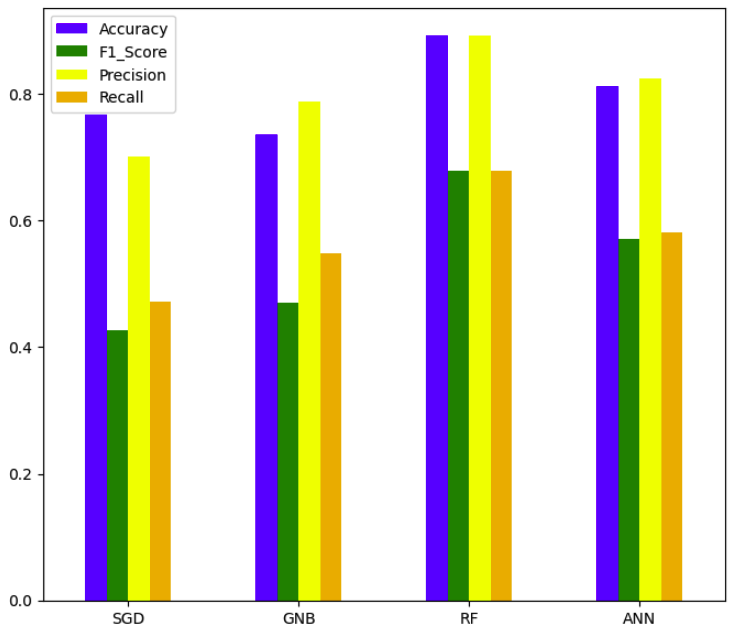}
    \caption{Reduced Feature Set}
    \label{fig: Select Feature Comparison}
  \end{subfigure}
  \hfill
  \begin{subfigure}[b]{0.51\textwidth}
    \includegraphics[width=\textwidth]{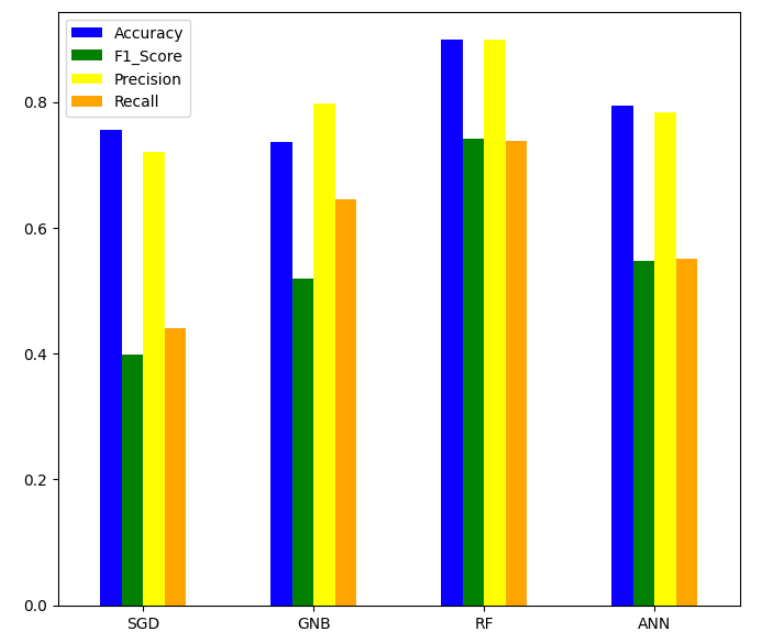}
    \caption{Full Feature Set}
    \label{fig: Full Feature Comparison}
  \end{subfigure}
\vspace{0.3cm}
  \caption{Select Feature and Full Feature Comparison.}
  \label{Model Comparisons}
\end{figure}

We compare the output of our model to several different machine learning models and contrast their performance to our artificial neural network (ANN). In particular, we review Random Forest Classification, Stochastic Gradient Descent, Gaussian Naive Bayes, and Support Vector Machines against our ANN model. We find Support Vector Machines to be inefficient for the size of our data set, taking too long to compute. However, amongst the other three models, we find Random Forest to be superior to our ANN model. To understand how our feature reduction compares (Figure:~\ref{fig: Select Feature Comparison}) to our full feature set (Figure:~\ref{fig: Full Feature Comparison}) with regards to impact on our model, we re-implement our Neural Network with all features. We also implement all features in our three comparison models as well.

%\vspace{-0.1cm}
%\vspace{0.2cm}

We find that our model performs with no significant difference between the full set of features, vs the reduced set of features. We also find that, compared to Random Forest Classification, our Neural network does not perform as well across all three evaluation metrics. 

\subsubsection{Refining our Neural Network Architecture}
Since our neural network model performed sub-optimally compared to our Random Forest Model, we look to improve our neural network architecture. Our initial implementation utilizes the Scikit package, which has limitations in designing an optimal neural network. We re-implement our neural network with TensorFlow. TensorFlow provides more control over the neural network architecture and allows us to tune the following parameters accordingly:

\setlength{\arrayrulewidth}{0.5mm}
\setlength{\tabcolsep}{10pt}
\renewcommand{\arraystretch}{1.5}
\begin{table*}[h!]
\centering
\begin{tabular}{ |p{5.5cm}|p{3cm}|p{4cm}| }
\hline
\multicolumn{3}{|c|}{\textbf{ANN Model Parameter Comparison}}\\
\hline
\textit{ANN Parameter} &\textit{Original Model (Scikit)} &\textit{Final Model  (Tensorflow)}\\
\hline
Loss Function  & Log Loss  & MAE (Mean Absolute Error) \\
Drop Out Rate & 0.0\% & 25.0\%\\
Layer 1 Activation Function & Relu & Relu\\
Layer 2 Activation Function  & Relu & Tanh\\
Output Layer Activation Function & Relu & Sigmoid \\
\hline
\end{tabular}
\caption{Refining our Model Architecture.}
\label{tab: Refining our Model Architecture}
\end{table*}

%\vspace{0.1cm}

\textbf{Loss Function}.
As our model only contains one output and that output is a corresponding value, rather than a pre-defined label, we select mean absolute error for our loss function over the log loss function that is intrinsically used within the Scikit library. Mean absolute error measures the variance in the distance between our predicted output in our model and the actual values in our test data set.

\textbf{Varying Activation Functions within our Layers}.
 We define an activation function for each hidden layer, vs a single activation function for all layers. We test different configurations of activation functions in each layer of our model. Our final model uses a different activation function at each layer. In between our input layer and our first hidden layer, we use a \textit{Relu} activation function; in between our first hidden layer and our second hidden layer, we use a \textit{Tanh} activation function; and in between our second hidden layer and our output layer, we use a \textit{Sigmoid} function. 

%ref Activation Functions, comparison in trends
%https://arxiv.org/pdf/1811.03378.pdf

\textbf{Utilizing Dropout for a More Precise Fit}.
Lastly to prevent overfitting our data, and to ensure our model more randomly fits our validation and test data, we implement a dropout layer of 25\% at each hidden layer.

\begin{figure}[ht!]
    \centering
    \includegraphics[width=0.9\textwidth, angle=0]{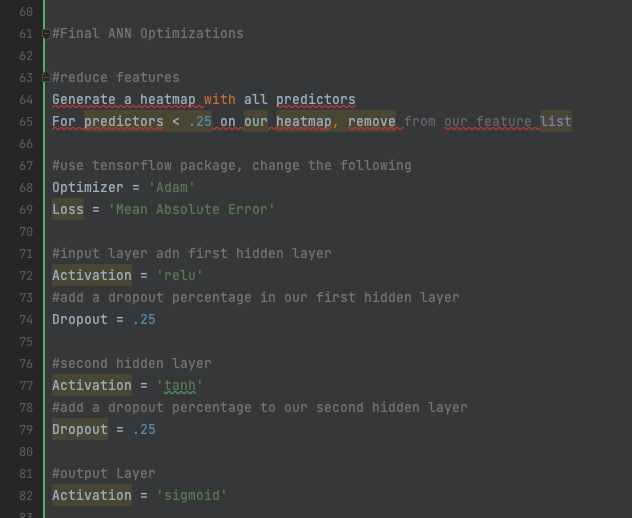}
    \caption{Final Model Improvements Pseudocode}
    \label{fig:Final Model Improvements Pseudocode}
\end{figure}

\textbf{Final Results for our Refined Neural Network Architecture}.
In our review, our revised architecture and final ANN model constructed with \textit{Tensorflow} outperformed both original ANN models with either \textit{select features} or the \textit{full set of features}. Our final ANN model \textit{also} performed comparably with either the select features or the full set of features. Review Table~\ref{tab: Comparing Neural Network Models}. This leads us to believe that a well-architected ANN model may be superior in performance and accuracy when compared to a restricted parameter and simple ANN model, and the ability to optimize computationally with only a select set of features did not negatively impact the results of the model in this particular case. 

%\vspace{1cm}
%\begin{figure*}[h!]
%\documentclass{article}
\setlength{\arrayrulewidth}{0.5mm}
\setlength{\tabcolsep}{10pt}
\renewcommand{\arraystretch}{1.5}
\begin{table*}[h!]
\centering
\begin{tabular}{|p{1.72cm}|p{3.8cm}|p{1.60cm}|p{1.60cm}|p{1.60cm}|p{1.5cm}|}
\hline
\multicolumn{6}{|c|}{\textbf{ANN Model Performance Comparison}}\\
\hline
\textit{Features}& \textit{ANN Model} &\textit{Accuracy} &\textit{F1 Score} &\textit{Precision}&\textit{Recall}\\
\hline
\multirow{2}{4em}{Select Features}& 
ANN Original Model & 0.8042 & 0.4853 & 0.8305 & 0.5905\\
& ANN Final    Model & 0.9506 & 0.9182 & 0.9394 & 0.9013\\
\hline
\multirow{2}{4em}{Full Features}&
ANN Original Model  & 0.7948 & 0.5470 & 0.7845 & 0.5511\\
& ANN Final    Model & 0.9443 & 0.9058 & 0.9278 & 0.8958\\
\hline
\end{tabular}
\caption{Comparing Neural Network Architecture and Accuracy}
\label{tab: Comparing Neural Network Models}
\end{table*}

%\end{figure*}
%\vspace{0.1cm}

\subsection{Discussion}

\textbf{The Effectiveness of Socioeconomic and Culturally Aware Features in Recommender Systems in Media Subscription Services}.
In this chapter, we reviewed the effectiveness of socioeconomic and culturally aware features with domain data to predict consumer preferences in price. This allowed us to understand the potential positive impact of socioeconomic and cultural features in use in recommender systems. In our motivating scenario, we found these features to positively impact price prediction with relative accuracy. We can extrapolate that predicting a consumer-initiated product movement is possible, such as a consumer either down-selecting or up-selecting a subscription service. 

This provides the potential for anticipating consumer movement in products, thus a company can possess
a deeper understanding of each customer. Anticipating an existing consumer's movement and preference allows for the development of preemptive custom interstitials that align with a consumer preference, therefore deepening the relationship between the customer and the company lending to better retention. For new customers, an enhanced design of a user experience can guide a user to a product selection that best anticipates their preference. 

\textbf{Generalizability in Markets beyond Media Subscription}.
The result of incorporating socioeconomic and cultural indices differentiate consumer from country to country and could allow businesses to suggest consumer changes as a way of retention when major socio-economic shifts occur either locally or globally.  It also seems plausible that a socio-economic and culturally aware recommender system could be used beyond determining subscription service preferences in the media industry, but apply to any subscription-based services, and more importantly be applied to any product pricing and price changes in other global industries where regional pricing may play a part. We conclude that a generalized model with socio-economic and cultural awareness of various different regions could be used to further personalize recommender systems where product pricing recommendations are made unique for various regional markets.

\textbf{Modeling Efficiency in Socio-economic and Culturally Aware Recommender Systems}.
Scientifically, and in terms of efficiently and effectively predicting consumer behavior in terms of price and plan selection with domain data, cultural indices, and socio-economic factors with a reasonable size data set, and minimal categorical data, we found that the most effective machine learning model was a well-architected artificial neural network model with \textit{select features} (Figure:~\ref{tab: Final ANN Model Performance Compared with other ML Models}. With this we conclude two specific thoughts with regard to recommender systems with similar data:  1) A well-architected neural network model with tested parameters may outperform traditional machine learning models where conditions are correct and computational limits are not imposed (e.g. GPU is available), 2) Random Forest machine learning models provide a reasonably efficient model with minimal tuning but may have a trade-off between precision and recall. We also learned that, with a well-architected neural network model, the feature set could be reduced, providing computational efficiency with similar accuracy.  

%\vspace{1cm}
\setlength{\arrayrulewidth}{0.5mm}
\setlength{\tabcolsep}{10pt}
\renewcommand{\arraystretch}{1.5}
\begin{table*}[h!]
\centering
\begin{tabular}{ |p{5.2cm}|p{1.8cm}|p{1.8cm}|p{1.8cm}|p{1.8cm}|}
\hline
\multicolumn{5}{|c|}{\textbf{Final Select Feature ML Model Performance Comparison}}\\
\hline
\textit{Model} &\textit{Accuracy} &\textit{F1 Score} &\textit{Precision} &\textit{Recall}\\
\hline
Stochastic Gradient Descent & 0.7666 & 0.4106 & 0.8994 & 0.4719\\
Gaussian Naive Bayes   & 0.7416 & 0.5046 & 0.8994 & 0.5489\\
Random Forest   & 0.8984 & 0.7146 & 0.8994 & 0.6781\\
ANN Final (Tensor Flow) & 0.9506 & 0.9182 & 0.9394 & 0.9013\\
\hline
\end{tabular}
\caption{Final ANN Model Performance Compared with other ML Models}
\label{tab: Final ANN Model Performance Compared with other ML Models}
\end{table*}

%%%%%%%%%%%%%%%%%%%%%%%%%%%%%%%%%%%%%%%%
\section{Conclusion and Future Work}

\subsection{Concluding Remarks}

With the consistent progression of more and more information and the limitations of human capability, methods of more accurately understanding a customer’s preference are essential. Recommender systems are necessary to understand a customer’s preference by automating processes that extract, sort, filter, and process information. Competition in the streaming media industry is gaining momentum. Globalization is a key to subscriber growth but adds complexities to differences and preferences in customers across varying regions. Customer preferences in different regions are impacted by environmental factors, such as those that are reflected in socio-economic and cultural norms. Therefore, incorporating socio-economic and cultural indicators as context to improve the accuracy of recommender systems is necessary.
 
In our research, we explored the predictability of socio-economic and cultural factors on a customer’s preference in plan selection.  We did this by predicting the price users in a particular country were willing to pay. We performed feature selection, fusing domain data and our socio-economic and cultural factors associated with the countries represented by our user base. We found that a well-architected artificial neural network with select cultural, and socio-economic features could predict the price customers were likely to want to pay with, with an accuracy of 95\%, a precision of 94\%, an F1 Score of 92\%, and a Recall of 90\%. 
It also performed more superior to other types of the other classification models we chose to experiment with, Random Forest, Gaussian Naive Bayes, and stochachastic gradient descent. Our well-architected ANN model uplifted our metrics across all evaluation dimensions by a minimum of 6\% in accuracy, 4\% in precision, 23\% in recall, and 20\% in F1, and did so with even greater consistency and balance (e.g. there is no notably significant trade-off between Precision and Recall). 

Therefore, the plan a consumer might select after an involuntary price change occurred is highly predictable, and this prediction could assist in the longer-term retention of customers by enabling a proactive recommendation to a customer whose intent is to leave the service. 

\subsection{Limitations}

 After the completion of our experiment, the research in this paper may present some limitations that in implementation could lead to varied results. 

\textbf{Limitations in Sample Data}.
We selected data from countries that we were able to more completely represent in our domain and non-domain data. However, certain geographical regions are absent, e.g., Africa, North Africa, Southeast Asia, and the Middle East. These regions may have more significant shifts in cultural and socioeconomic data, leading to potentially different features weighting in our model. 

\textbf{Training and Test Data Size}.
In order to narrow the size of our training and test data to match computational efficiency on a 32-core processor and 60gigabyte memory server, we narrowed the scope of our data. The scope was reduced to only those individuals who may react to involuntary price changes. In future research, we would run additional experiments across different profiles of users, not limited to only those that reacted to involuntary price changes. We may also correspondingly increase computing in order to increase computation efficiency and shorter durations for experiments. 

\subsection{Future Work}

There are different areas that we can propose to expand and improve on our research. Below are the main domains that we could explore in future works: 

\subsubsection{Temporal Drifts and Customer Behaviour} 

%Though cultural indicators rarely shift with time, socio-economic conditions due shift with time, and those shifts may impact shifts in customer preferences. Understanding customer behavior in relation to those socio-economic shifts may 1) enable more proactive pricing and plan selection, 2) unlock more accurate retention and subscriber growth forecasting models and 3) create more accurate pricing models per region.

In the context of temporal drifts and customer behavior, a promising avenue for future research involves examining the impact of socio-economic shifts on customer preferences. While cultural indicators tend to remain relatively stable over time, socio-economic conditions exhibit temporal variations that can influence changes in customer preferences. Understanding customer behavior in relation to these socio-economic shifts offers several potential benefits, including more proactive pricing and plan selection, improved retention and subscriber growth forecasting models, and the creation of more accurate pricing models per region.

To advance this field, future studies can focus on analyzing customer behavior patterns in conjunction with socio-economic shifts over time. By examining historical data and tracking changes in socio-economic indicators, such as income levels, employment rates, or inflation rates, it becomes possible to identify correlations between these shifts and changes in customer preferences.

Once the relationship between socio-economic shifts and customer behavior is established, future work can explore the integration of this knowledge into pricing and plan selection strategies. By proactively adapting pricing and plan offerings to align with anticipated customer preferences in response to socio-economic shifts, companies can enhance customer satisfaction and increase the likelihood of customer retention.

Furthermore, the understanding of customer behavior in relation to socio-economic shifts can also contribute to the development of more accurate forecasting models for retention and subscriber growth. By incorporating the temporal dynamics of socio-economic conditions, these models can provide better insights into future customer behavior and enable more precise predictions of customer retention and subscriber growth rates.

\subsubsection{Proactive Prediction of User’s Intent to Initiate Change}

%With retention as a key mechanism to sustaining growth, an additional area of research might explore the similarity between customer behaviors and proactively designed interstitials. These interstitials might suggest a recommendation prior to a customer-initiated request for action.

In the pursuit of sustainable growth through customer retention, an intriguing avenue for future research lies in exploring the similarity between customer behaviors and proactively designed interstitials. This approach aims to predict a user's intent to initiate change and recommend relevant suggestions even before the user explicitly requests action.
To advance this field, future studies can focus on analyzing user behaviors and identifying patterns that indicate an intent to initiate change. By examining various aspects of user interactions, such as browsing history, purchase patterns, or engagement metrics, it may be possible to identify behavioral signals that precede a user's desire for change.
Once these behavioral patterns are established, the next step involves designing interstitials that proactively suggest relevant recommendations to users. These interstitials, strategically placed within the user experience, can provide tailored suggestions or options that align with the anticipated change in user preferences or needs.
Also, advanced entity recognition and sentence ranking improve recommender systems' accuracy and help predict user intent, leading to more timely and relevant content suggestions~\cite{zhang2023multi}.

An essential aspect of this research would be to develop robust algorithms and methodologies that accurately predict user intent based on behavioral similarity. This may involve employing machine learning techniques, such as clustering or classification algorithms, to identify users exhibiting similar behavioural patterns. By leveraging these patterns, proactive predictions can be made regarding the users' intentions to initiate change.
Explainable AI can also enhance recommender systems by providing transparent, understandable reasons for suggestions, thereby increasing user trust, improving personalization, and aiding in regulatory compliance and system optimization~\cite{hanif2023comprehensive,hanif2022evidence}.
Reinforcement learning can also enhance recommender systems by continuously learning from user interactions to make increasingly personalized and effective recommendations, thereby improving user engagement and satisfaction~\cite{taheri2024partial}.
Another future line of work, can enhance deep reinforcement learning guided graph neural networks to revolutionize recommender systems by leveraging complex user-item interaction patterns and continuously learning from user feedback, leading to more sophisticated, adaptive, and effective recommendations~\cite{zhao2022deep}.

\subsubsection{Expanding Non-Domain Data}

Our research had fewer socioeconomic factors and did not describe any social media trends within those regions that may present relevant factors in media plan selection preferences. By extracting features in social media trends as additional socio-economic factors, we may be able to react more proactively and recommend plans to new users.  This may, therefore, positively impact subscriber growth in a media subscription-based company.

To advance this field, future studies can focus on identifying and extracting features from social media trends that are indicative of socioeconomic factors. By analyzing user-generated content, engagement patterns, and emerging topics on social media platforms, relevant insights can be gained regarding the preferences and behaviours of potential subscribers. These insights can then be incorporated as additional socioeconomic factors in the media plan selection process.

By expanding the range of socioeconomic factors considered, media subscription-based companies can make more informed and tailored recommendations to new users. This proactive approach enables the delivery of personalized plans that align with individual preferences and expectations~\cite{ghodratnama2023personalized}. By incorporating social media trends, which often reflect evolving societal interests and cultural shifts, the recommendations can stay abreast of the latest developments and ensure relevance to users' evolving preferences. Equity and regulation, however, may impact how personalized a model could be designed, and would need further review and investigation.

\subsubsection{Vector Based User Distances and Graphs}

%As there are sub-areas of our cultural and socio-economic based data, distinguishing this data would allow us to understand the distances and similarities between regions. To do this we may design data into separate vectors representing each subarea, and then graph the vectors using a queryable graph-based data store. These similarities might lead to better plans and price assignments or suggestions across countries.

In the domain of cultural and socio-economic data analysis, an intriguing direction for future research lies in harnessing the potential of vector-based representations and graph-based data stores to differentiate sub-areas within this data. This approach offers the opportunity to gain valuable insights into the distances and similarities between regions, ultimately leading to enhanced planning, price assignments, and recommendations across countries.

To advance this research direction, future studies can focus on designing data representations that capture the distinct sub-areas within cultural and socio-economic data. Each sub-area can be transformed into a separate vector representation, encapsulating the unique characteristics and attributes of that region. By employing appropriate feature extraction and vectorization techniques, the data can be effectively transformed into a format conducive to analysis and comparison.

Once the vectors representing different sub-areas are obtained, future work can explore the utilization of graph-based data stores for organizing and querying these vectors. Graph databases enable efficient storage and retrieval of complex interconnected data, allowing for flexible querying and analysis of relationships between the vectors. Constructing a queryable graph-based data store incorporating the vectors can facilitate the identification of similarities and distances between regions, thereby informing the development of enhanced recommendations and pricing strategies.

\subsubsection{Transfer Learning}

%The use of cultural and socio-economic data could be generalized and applied as context in recommender systems. Specifically, in actual media selection, such as TV series, socio-economic and cultural indicators may match specifically with users, and tailor more accurate viewing recommendations.

In the realm of recommender systems, an exciting avenue for future exploration lies in leveraging transfer learning techniques to enhance the incorporation of cultural and socio-economic data as contextual information. Specifically, this research direction holds promise in refining viewing recommendations in media selection domains, such as TV series, by aligning socio-economic and cultural indicators with individual users.

To advance this field, future studies can investigate the effectiveness of transfer learning approaches in capturing and transferring knowledge from diverse datasets containing socio-economic and cultural information. By leveraging pre-trained models or knowledge learned from similar domains, recommender systems can benefit from the transfer of relevant information to enhance their understanding of user preferences.

One potential avenue is to explore domain adaptation techniques in transfer learning, whereby models trained on a source domain, such as a dataset containing socio-economic and cultural indicators, are adapted to a target domain, such as a TV series recommendation dataset. This would enable the recommender system to effectively incorporate cultural and socio-economic context specific to each user, leading to more accurate and personalized viewing recommendations.

%%%%%%%%%%%%%%%%%%%%%%%%%%%%%%%%%%%%%%%%

\bibliographystyle{abbrv}
\bibliography{ms}

\begin{thebibliography}{10}

\bibitem{LinearTV2021Aug}
{Linear vs. Streaming: What{'}s Your Strategy?}, Aug. 2021.
\newblock [Online; accessed 3. Sep. 2022].

\bibitem{NetflixGlobal2016}
{Netflix Is Now Available Around the World - About Netflix}, Sept. 2022.
\newblock [Online; accessed 3. Sep. 2022].

\bibitem{Adomavicius2005}
G.~Adomavicius, R.~Sankaranarayanan, S.~Sen, and A.~Tuzhilin.
\newblock Incorporating contextual information in recommender systems using a
  multidimensional approach.
\newblock {\em ACM transactions on information systems}, 23(1):103--145, 2005.

\bibitem{Aggarwal2016}
C.~C. Aggarwal.
\newblock {\em Content-Based Recommender Systems}, pages 139--166.
\newblock Springer International Publishing, Cham, 2016.

\bibitem{AlanColman2014}
J.~H. Alan~Colman, Mahmoud~Hussein and M.~Kapuruge.
\newblock Context aware and adaptive systems.
\newblock {\em Context in Computing}, 2014.

\bibitem{barukh2021cognitive}
M.~C. Barukh, S.~Zamanirad, M.~Baez, A.~Beheshti, B.~Benatallah, F.~Casati,
  L.~Yao, Q.~Z. Sheng, and F.~Schiliro.
\newblock Cognitive augmentation in processes.
\newblock {\em Next-Gen Digital Services. A Retrospective and Roadmap for
  Service Computing of the Future: Essays Dedicated to Michael Papazoglou on
  the Occasion of His 65th Birthday and His Retirement}, pages 123--137, 2021.

\bibitem{beheshti2022knowledge}
A.~Beheshti.
\newblock Knowledge base 4.0: Using crowdsourcing services for mimicking the
  knowledge of domain experts.
\newblock In {\em 2022 IEEE International Conference on Web Services (ICWS)},
  pages 425--427. IEEE, 2022.

\bibitem{beheshti2023empowering}
A.~Beheshti.
\newblock Empowering generative ai with knowledge base 4.0: Towards linking
  analytical, cognitive, and generative intelligence.
\newblock In {\em 2023 IEEE International Conference on Web Services (ICWS)},
  pages 763--771. IEEE, 2023.

\bibitem{beheshti2017coredb}
A.~Beheshti, B.~Benatallah, R.~Nouri, V.~M. Chhieng, H.~Xiong, and X.~Zhao.
\newblock Coredb: a data lake service.
\newblock In {\em Proceedings of the 2017 ACM on Conference on Information and
  Knowledge Management}, pages 2451--2454, 2017.

\bibitem{beheshti2018corekg}
A.~Beheshti, B.~Benatallah, R.~Nouri, and A.~Tabebordbar.
\newblock Corekg: a knowledge lake service.
\newblock {\em Proceedings of the VLDB Endowment}, 11(12):1942--1945, 2018.

\bibitem{Beheshti2018a}
A.~Beheshti, B.~Benatallah, A.~Tabebordbar, H.~R. Motahari-Nezhad, M.~C.
  Barukh, and R.~Nouri.
\newblock {DataSynapse}: A social data curation foundry.
\newblock 37(3):351--384, 2018.

\bibitem{beheshti2020personality2vec}
A.~Beheshti, V.~Moraveji-Hashemi, S.~Yakhchi, H.~R. Motahari-Nezhad, S.~M.
  Ghafari, and J.~Yang.
\newblock personality2vec: Enabling the analysis of behavioral disorders in
  social networks.
\newblock In {\em Proceedings of the 13th international conference on web
  search and data mining}, pages 825--828, 2020.

\bibitem{beheshti2020towards}
A.~Beheshti, S.~Yakhchi, S.~Mousaeirad, S.~M. Ghafari, S.~R. Goluguri, and
  M.~A. Edrisi.
\newblock Towards cognitive recommender systems.
\newblock {\em Algorithms}, 13(8):176, 2020.

\bibitem{beheshti2023processgpt}
A.~Beheshti, J.~Yang, Q.~Z. Sheng, B.~Benatallah, F.~Casati, S.~Dustdar,
  H.~R.~M. Nezhad, X.~Zhang, and S.~Xue.
\newblock Processgpt: Transforming business process management with generative
  artificial intelligence.
\newblock {\em arXiv preprint arXiv:2306.01771}, 2023.

\bibitem{beheshti2016business}
S.-M.-R. Beheshti, B.~Benatallah, S.~Sakr, D.~Grigori, H.~R. Motahari-Nezhad,
  M.~C. Barukh, A.~Gater, and S.~H. Ryu.
\newblock Business process paradigms.
\newblock {\em Process analytics: Concepts and techniques for querying and
  analyzing process data}, pages 19--60, 2016.

\bibitem{BenSassi2017}
I.~{Ben Sassi}, S.~Mellouli, and S.~{Ben Yahia}.
\newblock Context-aware recommender systems in mobile environment: On the road
  of future research.
\newblock {\em Information Systems}, 72:27--61, 2017.

\bibitem{Bolton2003}
R.~N. Bolton and M.~B. Myers.
\newblock Price-based global market segmentation for services.
\newblock {\em Journal of marketing}, 67(3):108--128, 2003.

\bibitem{Bourdieu2013}
P.~Bourdieu.
\newblock {\em Distinction: A social critique of the judgement of taste}.
\newblock Routledge, 2013.

\bibitem{Braunhofer2011}
M.~Braunhofer, M.~Kaminskas, and F.~Ricci.
\newblock Recommending music for places of interest in a mobile travel guide.
\newblock In {\em Proceedings of the Fifth ACM Conference on Recommender
  Systems}, RecSys '11, page 253–256, New York, NY, USA, 2011. Association
  for Computing Machinery.

\bibitem{Buyrukoglu2022}
S.~Buyrukoğlu and A.~Akbaş.
\newblock Machine learning based early prediction of type 2 diabetes: A new
  hybrid feature selection approach using correlation matrix with heatmap and
  sfs.
\newblock {\em Balkan Journal of Electrical and Computer Engineering},
  10(2):110 -- 117, 2022.

\bibitem{Caprar2015}
D.~V. Caprar, T.~M. Devinney, B.~L. Kirkman, and P.~Caligiuri.
\newblock Conceptualizing and measuring culture in international business and
  management: From challenges to potential solutions.
\newblock {\em Journal of international business studies}, 46(9):1011--1027,
  2015.

\bibitem{Chaudhari2020}
K.~Chaudhari and A.~Thakkar.
\newblock A comprehensive survey on travel recommender systems.
\newblock {\em Archives of computational methods in engineering},
  27(5):1545--1571, 2020.

\bibitem{Chen2014}
L.~Chen and P.~Pu.
\newblock Experiments on user experiences with recommender interfaces.
\newblock {\em Behaviour and information technology}, 33(4):372--394, 2014.

\bibitem{Choi2014}
J.~Choi, H.~J. Lee, F.~Sajjad, and H.~Lee.
\newblock The influence of national culture on the attitude towards mobile
  recommender systems.
\newblock {\em Technological forecasting and social change}, 86(-):65--79,
  2014.

\bibitem{Chu2017}
W.-T. Chu and W.-H. Huang.
\newblock Cultural difference and visual information on hotel rating
  prediction.
\newblock {\em World wide web (Bussum)}, 20(4):595--619, 2017.

\bibitem{Dey2018}
S.~K. Dey, A.~Hossain, and M.~M. Rahman.
\newblock Implementation of a web application to predict diabetes disease: An
  approach using machine learning algorithm.
\newblock In {\em 2018 21st International Conference of Computer and
  Information Technology (ICCIT)}, pages 1--5, 2018.

\bibitem{Djordjevic2022}
K.~Djordjevic, M.~Jordovic~Pavlovic, Z.~Cojbasic, S.~Galovic, M.~Popovic,
  M.~Nesic, and D.~Markushev.
\newblock Influence of data scaling and normalization on overall neural network
  performances in photoacoustics.
\newblock {\em Optical and Quantum Electronics}, 54, 08 2022.

\bibitem{elahi2021recommender}
M.~Elahi, A.~Beheshti, and S.~R. Goluguri.
\newblock Recommender systems: Challenges and opportunities in the age of big
  data and artificial intelligence.
\newblock {\em Data Science and Its Applications}, pages 15--39, 2021.

\bibitem{ghodratnama2023personalized}
S.~Ghodratnama, A.~Behehsti, and M.~Zakershahrak.
\newblock A personalized reinforcement learning summarization service for
  learning structure from unstructured data.
\newblock In {\em 2023 IEEE International Conference on Web Services (ICWS)},
  pages 206--213. IEEE, 2023.

\bibitem{hanif2023comprehensive}
A.~Hanif, A.~Beheshti, B.~Benatallah, X.~Zhang, Habiba, E.~Foo, N.~Shabani, and
  M.~Shahabikargar.
\newblock A comprehensive survey of explainable artificial intelligence (xai)
  methods: Exploring transparency and interpretability.
\newblock In {\em International Conference on Web Information Systems
  Engineering}, pages 915--925. Springer, 2023.

\bibitem{hanif2022evidence}
A.~Hanif, A.~Beheshti, B.~Benatallah, X.~Zhang, and S.~Wood.
\newblock Evidence based pipeline for explaining artificial intelligence
  algorithms with interactions.
\newblock In {\em 2022 IEEE 9th International Conference on Data Science and
  Advanced Analytics (DSAA)}, pages 1--9. IEEE, 2022.

\bibitem{Haruna2017}
K.~Haruna, M.~Akmar~Ismail, S.~Suhendroyono, D.~Damiasih, A.~C. Pierewan,
  H.~Chiroma, and T.~Herawan.
\newblock Context-aware recommender system: A review of recent developmental
  process and future research direction.
\newblock {\em Applied Sciences}, 7(12), 2017.

\bibitem{he2023noise}
K.~He, S.~Meng, Q.~Li, X.~Liu, A.~Beheshti, X.~Chi, and X.~Zhang.
\newblock Noise-augmented contrastive learning for sequential recommendation.
\newblock In {\em International Conference on Web Information Systems
  Engineering}, pages 559--568. Springer, 2023.

\bibitem{Heaton2015}
J.~Heaton.
\newblock {\em Artificial Intelligence for Humans}, volume 3: Deep Learning and
  Neural Networks.
\newblock CreateSpace Independent Publishing Platform, 2015.

\bibitem{Hofstede2011}
G.~Hofstede.
\newblock Dimensionalizing cultures: The hofstede model in context.
\newblock {\em Online Readings in Psychology and Culture}, 2(1), 2011.

\bibitem{Hong2019}
M.~Hong, S.~An, R.~Akerkar, D.~Camacho, and J.~J. Jung.
\newblock Cross-cultural contextualisation for recommender systems.
\newblock {\em Journal of Ambient Intelligence and Humanized Computing}, 2019.

\bibitem{Hooijberg2007}
M.~Hooijberg.
\newblock Forward long line - vincenty's method.
\newblock In {\em Geometrical Geodesy}. Springer Berlin / Heidelberg, Germany,
  2007.

\bibitem{Huang2021}
L.~Huang, M.~Fu, F.~Li, H.~Qu, Y.~Liu, and W.~Chen.
\newblock A deep reinforcement learning based long-term recommender system.
\newblock {\em Knowledge-based systems}, 213:106706--, 2021.

\bibitem{Jabbar2014}
H.~Jabbar and R.~Z. Khan.
\newblock Methods to avoid over-fitting and under-fitting in supervised machine
  learning (comparative study).
\newblock {\em Computer Science, Communication and Instrumentation Devices},
  70:163--172, 2015.

\bibitem{jalayer2022ham}
A.~Jalayer, M.~Kahani, A.~Pourmasoumi, and A.~Beheshti.
\newblock Ham-net: Predictive business process monitoring with a hierarchical
  attention mechanism.
\newblock {\em Knowledge-Based Systems}, 236:107722, 2022.

\bibitem{Jannach2016}
D.~Jannach, P.~Resnick, A.~Tuzhilin, and M.~Zanker.
\newblock Recommender systems — beyond matrix completion.
\newblock {\em Commun. ACM}, 59(11):94–102, oct 2016.

\bibitem{Jones2014}
N.~Jones.
\newblock The learning machines.
\newblock {\em Computer science: The learning machines}, 505(7482):146–148,,
  2014.

\bibitem{Jun2020}
H.~J. Jun, J.~H. Kim, D.~Y. Rhee, and S.~W. Chang.
\newblock "seoulhouse2vec": An embedding-based collaborative filtering housing
  recommender system for analyzing housing preference.
\newblock {\em Sustainability (Basel, Switzerland)}, 12(17):6964--, 2020.

\bibitem{Kalinowski2021}
K.~Kalinowski, A.~Kozłowska, M.~Malesza, and D.~P. Danel.
\newblock Evolutionary origins of music. classical and recent hypotheses.
\newblock {\em Anthropological review (Poznań, Poland)}, 84(2):213--231, 2021.

\bibitem{Kaminskas2012}
M.~Kaminskas and F.~Ricci.
\newblock Contextual music information retrieval and recommendation: State of
  the art and challenges.
\newblock {\em Computer Science Review}, 6(2):89--119, 2012.

\bibitem{Karlik2011}
B.~Karlik and A.~Vehbi.
\newblock Performance analysis of various activation functions in generalized
  mlp architectures of neural networks.
\newblock {\em International Journal of Artificial Intelligence and Expert
  Systems (IJAE)}, 1(4):pp. 111–122, 2011.

\bibitem{khadivizand2020towards}
S.~Khadivizand, A.~Beheshti, F.~Sobhanmanesh, Q.~Z. Sheng, E.~Istanbouli,
  S.~Wood, and D.~Pezaro.
\newblock Towards intelligent feature engineering for risk-based customer
  segmentation in banking.
\newblock In {\em Proceedings of the 18th International Conference on Advances
  in Mobile Computing \& Multimedia}, pages 74--83, 2020.

\bibitem{Kingma2017}
D.~P. Kingma and J.~Ba.
\newblock Adam: A method for stochastic optimization.
\newblock In {\em 3rd International Conference for Learning Representations},
  2017.

\bibitem{Koenigstein2011}
N.~Koenigstein, G.~Dror, and Y.~Koren.
\newblock Yahoo! music recommendations: modeling music ratings with temporal
  dynamics and item taxonomy.
\newblock In {\em Proceedings of the fifth ACM conference on recommender
  systems}, RecSys '11, pages 165--172. ACM, 2011.

\bibitem{Koren2009}
Y.~Koren.
\newblock Collaborative filtering with temporal dynamics.
\newblock In {\em Proceedings of the 15th ACM SIGKDD international conference
  on knowledge discovery and data mining}, KDD '09, pages 447--456. ACM, 2019.

\bibitem{Kraemer2005}
K.~L. Kraemer, J.~Gibbs, and J.~Dedrick.
\newblock Impacts of globalization on e-commerce use and firm performance: A
  cross-country investigation.
\newblock {\em The Information Society}, 21(5):323--340, 2005.

\bibitem{Kweon2021}
H.~J. Kweon and S.~H. Kweon.
\newblock Pricing strategy within the u.s. streaming services market: A focus
  on netflix's price plans.
\newblock {\em International journal of contents}, 17(2):1--8, 2021.

\bibitem{li2022survey}
Q.~Li, J.~Li, J.~Sheng, S.~Cui, J.~Wu, Y.~Hei, H.~Peng, S.~Guo, L.~Wang,
  A.~Beheshti, et~al.
\newblock A survey on deep learning event extraction: Approaches and
  applications.
\newblock {\em IEEE Transactions on Neural Networks and Learning Systems},
  2022.

\bibitem{li2021comprehensive}
Q.~Li, H.~Peng, J.~Li, Y.~Hei, R.~Sun, J.~Sheng, S.~Guo, L.~Wang, J.~Wu,
  A.~Beheshti, et~al.
\newblock A comprehensive survey on schema-based event extraction with deep
  learning.
\newblock {\em arXiv preprint arXiv:2107.02126}, 2021.

\bibitem{Liang2022}
J.~Liang.
\newblock Confusion matrix: Machine learning.
\newblock {\em POGIL Activity Clearinghouse}, 3(4), Dec. 2022.

\bibitem{liu2022dagad}
F.~Liu, X.~Ma, J.~Wu, J.~Yang, S.~Xue, A.~Beheshti, C.~Zhou, H.~Peng, Q.~Z.
  Sheng, and C.~C. Aggarwal.
\newblock Dagad: Data augmentation for graph anomaly detection.
\newblock In {\em 2022 IEEE International Conference on Data Mining (ICDM)},
  pages 259--268. IEEE, 2022.

\bibitem{Liu2018}
M.~Liu, X.~Hu, and M.~Schedl.
\newblock The relation of culture, socio-economics, and friendship to music
  preferences: A large-scale, cross-country study.
\newblock {\em PloS one}, 13(12):e0208186--e0208186, 2018.

\bibitem{Moradizeyveh2022intent}
S.~Moradizeyveh.
\newblock Intent recognition in conversational recommender systems.
\newblock {\em arXiv preprint arXiv:2212.03721}, 2022.

\bibitem{Nwankpa2018}
C.~Nwankpa, W.~Ijomah, A.~Gachagan, and S.~Marshall.
\newblock Activation functions: Comparison of trends in practice and research
  for deep learning.
\newblock {\em arXiv preprint arXiv:1811.03378}, 2018.

\bibitem{PatrickBrezillon2014}
A.~J.~G. Patrick~Brezillon.
\newblock Preface by the editors.
\newblock {\em Context in Computing}, 2014.

\bibitem{sensemaking}
P.~Pirolli and S.~Card.
\newblock The sensemaking process and leverage points for analyst technology as
  identified through cognitive task analysis.
\newblock In {\em Proceedings of international conference on intelligence
  analysis}, volume~5, pages 2--4. McLean, VA, USA, 2005.

\bibitem{K.He2015}
K.~H. Z.~S. Ren; and J.~Sun.
\newblock Delving deep into rectifiers: Surpassing human-level performance on
  imagenet classification.
\newblock {\em arXiv}, 2015.

\bibitem{Ricci2022}
F.~Ricci.
\newblock {\em Recommender Systems Handbook.}
\newblock Springer, New York, NY, 3rd ed. edition, 2022.

\bibitem{Rosa2019}
R.~L. Rosa, G.~M. Schwartz, W.~V. Ruggiero, and D.~Z. Rodriguez.
\newblock A knowledge-based recommendation system that includes sentiment
  analysis and deep learning.
\newblock {\em IEEE transactions on industrial informatics}, 15(4):2124--2135,
  2019.

\bibitem{rouhollahi2021towards}
Z.~Rouhollahi, A.~Beheshti, S.~Mousaeirad, and S.~R. Goluguri.
\newblock Towards proactive financial crime and fraud detectionthrough
  artificial intelligence and regtech technologies.
\newblock In {\em The 23rd International Conference on Information Integration
  and Web Intelligence}, pages 538--546, 2021.

\bibitem{Schedl2021}
M.~Schedl, C.~Bauer, W.~Reisinger, D.~Kowald, and E.~Lex.
\newblock Listener modeling and context-aware music recommendation based on
  country archetypes.
\newblock {\em Frontiers in Artificial Intelligence}, 3, 2021.

\bibitem{Scheel2014}
C.~Scheel, A.~Castellanos, T.~Lee, and E.~W. De~Luca.
\newblock The reason why: A survey of explanations for recommender systems.
\newblock In {\em Adaptive Multimedia Retrieval: Semantics, Context, and
  Adaptation}, volume 8382 of {\em Lecture Notes in Computer Science}, pages
  67--84, Cham, 2014. Springer International Publishing.

\bibitem{schiliro2022deepcog}
F.~Schiliro, N.~Moustafa, I.~Razzak, and A.~Beheshti.
\newblock Deepcog: A trustworthy deep learning-based human cognitive privacy
  framework in industrial policing.
\newblock {\em IEEE Transactions on Intelligent Transportation Systems}, 2022.

\bibitem{shabani2023rule}
N.~Shabani, A.~Beheshti, H.~Farhood, M.~Bower, M.~Garrett, and
  H.~Alinejad-Rokny.
\newblock A rule-based approach for mining creative thinking patterns from big
  educational data.
\newblock {\em AppliedMath}, 3(1):243--267, 2023.

\bibitem{shabani2023survey}
N.~Shabani, J.~Wu, A.~Beheshti, J.~Foo, A.~Hanif, and M.~Shahabikargar.
\newblock A survey on graph neural networks for graph summarization.
\newblock {\em arXiv preprint arXiv:2302.06114}, 2023.

\bibitem{Shi2014}
Y.~Shi, M.~Larson, and A.~Hanjalic.
\newblock Collaborative filtering beyond the user-item matrix: A survey of the
  state of the art and future challenges.
\newblock {\em ACM Comput. Surv.}, 47(1), may 2014.

\bibitem{sobhanmanesh2023cognitive}
F.~Sobhanmanesh, A.~Beheshti, N.~Nouri, N.~M. Chapparo, S.~Raj, and R.~A.
  George.
\newblock A cognitive model for technology adoption.
\newblock {\em Algorithms}, 16(3):155, 2023.

\bibitem{NatGeographicEnc}
N.~G. Society.
\newblock Globalization.

\bibitem{Srivastava2014}
N.~Srivastava, G.~Hinton, A.~Krizhevsky, I.~Sutskever, and R.~Salakhutdinov.
\newblock Dropout: A simple way to prevent neural networks fromoverfitting.
\newblock {\em Journal of Machine Learning}, 15:1929, 2014.

\bibitem{tabebordbar2019}
A.~Tabebordbar, A.~Beheshti, and B.~Benatallah.
\newblock {ConceptMap}: A conceptual approach for formulating user preferences
  in large information spaces.
\newblock In {\em Web Information Systems Engineering {\textendash} {WISE}
  2019}, Lecture Notes in Computer Science, pages 779--794. Springer
  International Publishing, 2019.

\bibitem{taheri2024partial}
A.~Taheri, K.~RahimiZadeh, A.~Beheshti, J.~Baumbach, R.~V. Rao, S.~Mirjalili,
  and A.~H. Gandomi.
\newblock Partial reinforcement optimizer: An evolutionary optimization
  algorithm.
\newblock {\em Expert Systems with Applications}, 238:122070, 2024.

\bibitem{Trompenaars1998}
F.~Trompenaars and C.~Hampden-Turner.
\newblock {\em Riding the waves of culture: Understanding diversity in global
  business}.
\newblock McGraw Hill, 1998.

\bibitem{wang2023learning}
S.~Wang, A.~Beheshti, Y.~Wang, J.~Lu, Q.~Z. Sheng, S.~Elbourn, and
  H.~Alinejad-Rokny.
\newblock Learning distributed representations and deep embedded clustering of
  texts.
\newblock {\em Algorithms}, 16(3):158, 2023.

\bibitem{YakhchiShahpar2021LCUP}
S.~Yakhchi.
\newblock Learning complex users' preferences for recommender systems.
\newblock {\em arXiv preprint arXiv:2107.01529}, 2021.

\bibitem{yakhchi2022convolutional}
S.~Yakhchi, A.~Behehsti, S.-m. Ghafari, I.~Razzak, M.~Orgun, and M.~Elahi.
\newblock A convolutional attention network for unifying general and sequential
  recommenders.
\newblock {\em Information Processing \& Management}, 59(1):102755, 2022.

\bibitem{yakhchi2020towards}
S.~Yakhchi, A.~Beheshti, S.-M. Ghafari, M.~A. Orgun, and G.~Liu.
\newblock Towards a deep attention-based sequential recommender system.
\newblock {\em IEEE Access}, 8:178073--178084, 2020.

\bibitem{Ying2019}
X.~Ying.
\newblock An overview of overfitting and its solutions.
\newblock {\em Journal of Physics: Conference Series}, 1168(2):022022, feb
  2019.

\bibitem{Zafari2019}
F.~Zafari, I.~Moser, and T.~Baarslag.
\newblock Modelling and analysis of temporal preference drifts using a
  component-based factorised latent approach.
\newblock {\em Expert systems with applications}, 116:186--208, 2019.

\bibitem{Zangerle2018}
E.~Zangerle, M.~Pichl, and M.~Schedl.
\newblock Culture-aware music recommendation.
\newblock In {\em Proceedings of the 26th Conference on user modeling,
  adaptation and personalization}, UMAP '18, pages 357--358. ACM, 2018.

\bibitem{zhang2023multi}
G.~Zhang, X.~Zhang, Z.~Wang, and A.~Beheshti.
\newblock Multi-granularity entity recognition based sentence ranking for
  multi-document summarization.
\newblock In {\em 2023 IEEE 10th International Conference on Data Science and
  Advanced Analytics (DSAA)}, pages 1--10. IEEE, 2023.

\bibitem{zhao2022deep}
X.~Zhao, J.~Wu, H.~Peng, A.~Beheshti, J.~J. Monaghan, D.~McAlpine,
  H.~Hernandez-Perez, M.~Dras, Q.~Dai, Y.~Li, et~al.
\newblock Deep reinforcement learning guided graph neural networks for brain
  network analysis.
\newblock {\em Neural Networks}, 154:56--67, 2022.

\bibitem{Mowafi2014}
Y.~M. R. A.~A. Zmily.
\newblock User-centered approaches to context awareness: Prospects and
  challenges.
\newblock {\em Context in Computing}, 2014.

\end{thebibliography}

\end{document}